\documentclass[twocolumn]{aastex631}
\usepackage{threeparttable}
\usepackage{amssymb}
\usepackage{graphicx}
\usepackage[whole]{bxcjkjatype}
\usepackage{natbib,textcomp}
\usepackage{gensymb,amsmath}
\usepackage[caption=false]{subfig}

\newcommand{\dn}{$\mbox{D}_n\mbox{4000}$}
\newcommand{\ewhd}{$\mbox{EW(H}\delta\mbox{)}$}
\newcommand{\kms}{km~s$^{-1}$}
\newcommand{\othreea}{[\ion{O}{3}]$\lambda$5007}
\newcommand{\othreeb}{[\ion{O}{3}]$\lambda$4959}

\begin{document}

\title{Stars, gas, and star formation of distant post-starburst galaxies}

	\author{Po-Feng Wu \begin{CJK*}{UTF8}{bkai}(吳柏鋒)\end{CJK*}}
	\affiliation{Institute of Astrophysics, National Taiwan University, Taipei 10617, Taiwan}
	\affiliation{Department of Physics and Center for Theoretical Physics, National Taiwan University, Taipei 10617, Taiwan}
	\affiliation{Physics Division, National Center for Theoretical Sciences, Taipei 10617, Taiwan}

        \author{Rachel Bezanson}
        \affiliation{Department of Physics and Astronomy and PITT PACC, University of Pittsburgh, Pittsburgh, PA 15260, USA}

        \author{Francesco D'Eugenio}
        \affiliation{Kavli Institute for Cosmology, University of Cambridge, Madingley Road, Cambridge, CB3 0HA, United Kingdom}
        \affiliation{Cavendish Laboratory - Astrophysics Group, University of Cambridge, 19 JJ Thomson Avenue, Cambridge, CB3 0HE, United Kingdom}

        \author{Anna R. Gallazzi}
        \affiliation{INAF-Osservatorio Astrofisico di Arcetri, Largo Enrico Fermi 5, 50125 Firenze, Italy}

        \author{Jenny E. Greene}
        \affiliation{Department of Astrophysical Sciences, Princeton University, Princeton, NJ 08544, USA}

        \author{Michael V. Maseda}
        \affiliation{Department of Astronomy, University of Wisconsin-Madison, 475 N. Charter St., Madison, WI 53706, USA}

        \author{Katherine A. Suess}
        \affiliation{Department of Astronomy and Astrophysics, University of California, Santa Cruz, 1156 High Street, Santa Cruz, CA 95064, USA}
        \affiliation{Kavli Institute for Particle Astrophysics and Cosmology and Department of Physics, Stanford University, Stanford, CA 94305, USA}

        \author{Arjen van der Wel}
        \affiliation{Sterrenkundig Observatorium, Universiteit Gent, Krijgslaan 281 S9, B-9000 Gent, Belgium}

	\correspondingauthor{Po-Feng Wu}
	\email{wupofeng@phys.ntu.edu.tw}

\begin{abstract}

We present a comprehensive multi-wavelength study of 5 poststarburst galaxies with $M_\ast > 10^{11} M_\odot$ at $z\sim 0.7$, examining their stars, gas, and current and past star-formation activities. Using optical images from the Subaru telescope and Hubble Space Telescope, we observe a high incidence of companion galaxies and low surface brightness tidal features, indicating that quenching is closely related to interactions between galaxies. From optical spectra provided by the LEGA-C survey, we model the stellar continuum to derive the star-formation histories and show that the stellar masses of progenitors ranging from $2\times10^9 M_\odot$ to $10^{11} M_\odot$, undergoing a burst of star formation several hundred million years prior to observation, with a decay time scale of $\sim100$ million years. Our ALMA observations detect CO(2-1) emission in four galaxies, with the molecular gas spreading over up to $>1"$, or $\sim10$ kpc, with a mass of up to $\sim2 \times10^{10} M_\odot$. However, star-forming regions are unresolved by either the slit spectra or 3~GHz continuum observed by the Very Large Array. Comparisons between the star-formation rates and gas masses, and the sizes of CO emission and star-forming regions suggest a low star-forming efficiency. We show that the star-formation rates derived from IR and radio luminosities with commonly-used calibrations tend to overestimate the true values because of the prodigious amount of radiation from old stars and the contribution from AGN, as the optical spectra reveal weak AGN-driven outflows.

\end{abstract}


\section{Introduction}

Over the past few decades, extensive extragalactic surveys have unveiled the existence of two main categories of galaxies: the star-forming galaxies, which are actively undergoing in-situ star-formation, and the quiescent galaxies, whose star-formation rates (SFRs) are much lower compared to their stellar mass budgets \citep{str01,wil06,fra07}. Investigations into the early Universe have shown that the number density and total stellar mass density of quiescent galaxies increase with time since $z\sim4$ \citep{bel04,fab07,ilb13,muz13a}. Given the low levels of star formation in quiescent galaxies, the growth in their mass budgets and number demands that a portion of the star-forming galaxies must have ceased star formation and joined the quiescent population at any given time.

Observations have characterized large samples of transitioning galaxies and have found that galaxies can spend varying lengths of time moving from the star-forming to the quiescent category, driven by multiple mechanisms \citep{mar07,barr13,sch14,wu18a,bel19}. Some galaxies undergo rapid truncation of star formation, known as quenching, which contributes a smaller fraction of the quiescent population in the local Universe but is likely to be the dominant process at $z\gtrsim1$ \citep{gon12,wil16,row18}. Immediately following the sudden halt of star formation, the spectral energy distributions (SEDs) of galaxies are briefly dominated by A-type stars. These galaxies are referred to as \textit{poststarburst galaxies} \citep{dre83} and offer valuable insights into the physical conditions near the time of quenching.

The quest to understand the suppression of star formation in the early Universe through the examination of high-z poststarburst galaxies is both illuminating and challenging. This phase is expected to last for just a few hundred million years, much shorter than the lifetimes of galaxies, hence poststarburst galaxies are rare objects \citep{leb06,sny11}. Their defining characteristic, a dominant A-type stellar population, is difficult to identify without either multiwavelength photometry from restframe UV to near IR or deep spectra covering key diagnostic spectral lines \citep{bal99,dre99,kri11,whi12,wu20}, both of which are resource-intensive. Furthermore, with low SFRs by definition, poststarburst galaxies are likely to exhibit weak optical emission lines, far-infrared continuum emission, and low cold gas contents, making the observation of these properties beyond the local Universe a costly endeavor.

Despite the cost, the study of distant poststarburst galaxies has been gradually accumulating knowledge. With the aid of sophisticated algorithms and stellar population synthesis models, high signal-to-noise optical spectra can not only identify poststarburst galaxies but also constrain critical parameters of their formation histories such as the timing and intensity of the preceding burst events prior to quenching \citep{wil20,sue22}. Meanwhile, deep and high-resolution optical images reveal that poststarburst galaxies tend to be more compact than the typical quiescent galaxies, yet often display disturbed morphologies \citep{wil09,ver10,wu14,wu18a,set22}. This points towards a formation scenario in which quenching in distant Universe is the result of intense central starbursts brought on by gravitational interactions between galaxies \citep{wu20,him23}. However, despite such findings, deep rest-frame optical spectroscopy that is resolved spatially does not always find positive stellar age gradients, as one would expect if the central starbursts had happened \citep{deu20a,set20}.

The complexity of the problem of poststarburst galaxy studies is compounded by the presence of dust. It is well known that dust forms in tandem with the intense starburst event, thus, the young stellar population may be obscured and its light highly attenuated or even completely shielded \citep{pog00,bar22,sme22}. This presents a challenge in that the optical light may no longer accurately reflect the true nature of the stellar population. The emission from dust is also problematic, as many poststarburst galaxies selected at $z\sim1$ by the absence of emission lines exhibit a non-negligible amount of mid-IR emission \citep{koc11,wu14}. This inconsistency between star-formation tracers has yet to be resolved.

In terms of the raw material for star-formation, molecular gas, the Atacama Large Millimeter/submillimeter Array (ALMA) has revealed that some distant massive poststarburst galaxies contain molecular gas amounting to more than $10\%$ of their stellar mass contents \citep{bez22}. This finding raises questions about the mechanisms that must be in operation to account for the lack of star formation despite the availability of fuel. Although \ion{Mg}{2}$\lambda2796,2803$ outflows at velocities of $\sim 1000$ \kms\ have been discovered in several cases \citep{tre07,mal18}, there is yet to be clear evidence of molecular outflow in distant poststarburst galaxies. The hypothesis that active galactic nuclei (AGNs) play a significant role in the quenching process has been suggested \citep{spr05,som15}, but clear signs of AGN such as point-source X-ray or radio continuum emission do not always present in distant poststarburst galaxies \citep{ver10}. Deep rest-frame optical spectra have shown that the nebular emission lines in poststarburst galaxies are more often driven by low-ionization nuclear emission-line regions (LINERs) rather than AGNs \citep{lem10,lem17}.

Our knowledge of quenching processes beyond the local Universe has been steadily growing through the accumulation of observations. Nevertheless, it has been challenging to synthesize the findings due to the different galaxy samples and selection criteria used by various studies. Each study is limited by the available data and adopts different selection methods, picking up intrinsically different samples in terms of stellar ages, AGN, and etc \citep{yes14,wu20,bez22}. It is thus not straightforward to assume the properties observed in one sample can apply to another.

In this paper, we present a comprehensive analysis of 5 poststarburst galaxies at $z\sim0.7$. The dataset includes ultra-deep optical slit spectra ($S/N > 20\ \AA^{-1}$), interferometric observations of CO(2-1) molecular line, optical images from the Hubble Space Telescope (HST), as well as deep multi-wavelength images from X-ray to radio. This extensive dataset enables us to examine the physical properties of stars, gas, and star-formation activities from the same set of galaxies, providing a self-consistent understanding of the quenching process in the distant Universe.

We will explore various hypotheses surrounding the quenching process that occurred approximately 7 billion years ago. These hypotheses include whether galaxy-galaxy interactions triggered the quenching, whether a centrally-concentrated starburst occurred prior to the quenching, and the role of AGN in the gas content and quenching. Additionally, we will investigate the accuracy of star-formation rate (SFR) tracers at different wavelengths, as this information is crucial for determining the star-formation efficiency (SFE), i.e., the ratio of SFR to gas mass.

We detail our galaxy sample, data, and methodology in Section~\ref{sec:data}. We present the spatial distributions and kinematics of stars, gas, and star-formation in Section~\ref{sec:dist} and Section~\ref{sec:kin}. The SFRs, molecular gas contents, and their relations are discussed in Section~\ref{sec:sfr} and Section~\ref{sec:gas}. The implications of these results for quenching are discussed in Section~\ref{sec:dis}. Finally, our summary and conclusions can be found in Section~\ref{sec:sum}.

\section{Data and Analysis}
\label{sec:data}

Poststarburst galaxies in this work are selected based on their spectra taken by the Large Extragalactic Galaxy Astrophysics Census survey \citep[LEGA-C,][]{vdw16} and then follow-up by ALMA to measure the CO(2-1) emission. All galaxies are in the COSMOS field \citep{sco07}, and plenty of ancillary multi-wavelength data are publicly available. 

\subsection{The LEGA-C survey and sample selection}
\label{sec:sample}

\begin{figure}
    \includegraphics[width=0.95\columnwidth]{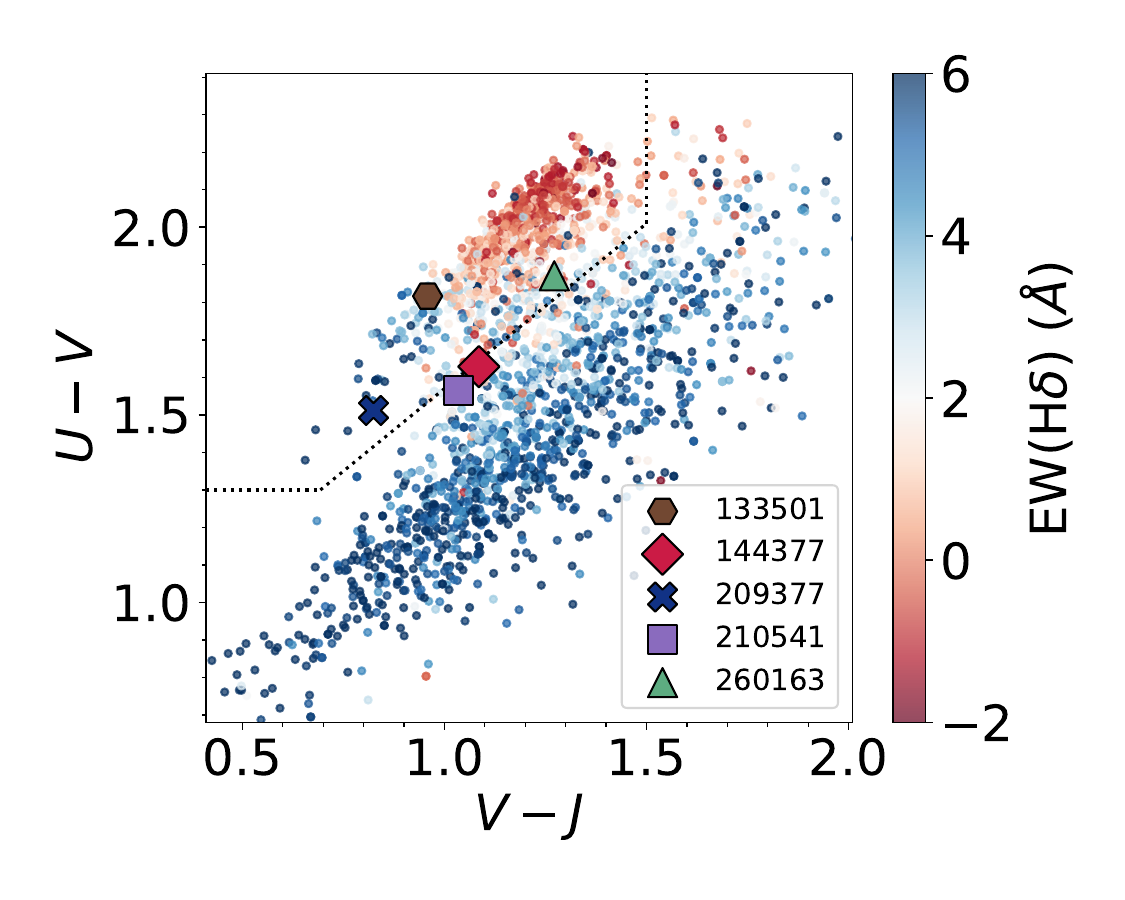}
    \caption{The rest-frame U-V and V-J color-color diagram of poststarburst galaxies in this study. Small dots are galaxies in the LEGA-C survey with $0.6 \leq z \leq 1.0$, color-coded by their \ewhd. The dotted lines separate quiescent galaxies at the upper left from star-forming galaxies at the bottom right. Post-starburst galaxies, i.e., quiescent galaxies with strong Balmer absorption, tend to, but not exclusively be with the bluest V-J colors. Spectroscopy can pick up poststarburst galaxies that would not be identified with photometries. \label{fig:uvj}  }
\end{figure}

In recent years, a variety of methods have been proposed and adopted to identify galaxies with recently rapidly declining SFRs. Each selection method carries its own limitations and biases \citep[e.g., see discussion in][]{yes14,wu20}. The selection criteria should depend on the data available and scientific goals. The LEGA-C survey obtains rest-frame optical spectra of $\sim3000$ galaxies at $0.6\leq z \leq1.0$. The typical continuum signal-to-noise ratio (S/N) is 20 \AA$^{-1}$ with a spectral resolution of $R\simeq3500$ \citep{str18,vdw21}, allowing us to use spectral features to select the sample as in many early studies \citep[e.g.][]{zab96,got05}.

Poststarburst galaxies are often identified as those galaxies with strong Balmer absorption and weak emission lines. We use the H$\delta$ absorption line as the tracer for the star-formation activities in the recent past. For the star-formation tracer, we choose H$\beta$ instead of the commonly-used [\ion{O}{2}]$\lambda3727,3729$ doublet because the [\ion{O}{2}] emission in galaxies with low SFRs are likely dominated by AGN or old stars \citep{lem10,lem17,mas21}.

We define poststarburst galaxies as those with
\begin{equation} \label{eq:psb}
	\begin{aligned}
		EW(H\beta)_{em} & \geq -1.5\mbox{\AA} \\
		EW(H\delta)  & \geq 4\mbox{\AA}. \\
	\end{aligned}
\end{equation}

as measured from the 2nd data release \citep[DR2,][]{str18} of the LEGA-C project. \citet{wu18a} showed that $EW(H\beta)_{em} = -1\mbox{\AA}$ separates star-forming from quiescent galaxies well. Here we adopt a slightly higher equivalent width cut to potentially include galaxies with poststarburst-like SFHs but slightly higher SFRs. These looser definitions do not lead to a contaminated sample, as we will show later in the paper.

To measure the line strengths, we employ a joint fitting of both the stellar continuum and emission lines using the Penalized Pixel-Fitting method \citep[pPFX;][]{cap04,cap17}, as described in \citep{str18}. We use the Lick H$\delta_a$ definition \citep{wor94} to measure the equivalent width of H$\delta$ and determine it from the spectra after subtracting the emission line models generated by \texttt{pPFX}.
In DR2, the emission-line fluxes are directly taken from the results of \texttt{pPXF} fit and EWs were derived by dividing the line fluxes by the local continuum averaging over a 200\AA-wide region centered on the line.

For ALMA follow-ups, we select galaxies at redshifts $z<0.84$, as the CO(2-1) line can be observed with ALMA Band 4 receivers. A lower mass limit of $\log(M_\ast/M_\odot)>10.8$ is also set. We further require galaxies to have either position angles $P.A. < 20\degree$ or axis ratios $b/a > 0.7$ measured from \textit{HST} F814W images. As LEGA-C slits are oriented north-south, these galaxies are easier to be spatially resolved in the slit spectra.

From 1550 primary targets in LEGA-C DR2, spectra of 678 galaxies cover both  $EW(H\beta)_{em}$ and $EW(H\delta)$. Thirty-seven galaxies fulfill the equivalent width cut of Equation~\ref{eq:psb} and 6 of them also meet our criteria of redshift, stellar mass, and shapes. All 6 of them are proposed for ALMA observation and 5 are observed.

Table~\ref{tab:sample} lists the spectral features of the 5 galaxies. We use DR2 spectra for sample selection but the analysis is done with the 3rd data release \citep[DR3,][]{vdw21}. We thus list measurements from both sets. We also list the widely-used spectral index \dn. The DR3 improves sky subtraction and flux calibration, therefore, the overall shapes of spectra and the line strengths may change between DR2 and DR3. Furthermore, the emission line fluxes and equivalents widths are measured differently in DR3. The DR3 uses \texttt{Platefit} \citep{tre04,bri04} on spectra after subtracting the best-fit continuum model to fit the emission lines. The equivalent width of the emission line is calculated based on the measured line flux and the continuum flux from the continuum model over the same wavelength range \citep{mas21}. The reason for not using \texttt{pPXF} for emission-line fluxes is that \texttt{pPXF} does not measure faint lines well and \texttt{Platefit} is customized for such a task.

Between DR3 and DR2, the scatters in \dn\ and \ewhd\ are 0.08 and 0.8\AA, respectively. For EW(H$\beta$)$_{em}$, the DR3 has $\sim20\%$ more detections above 3$\sigma$ uncertainties than DR2 and DR3 is on average $\sim1.3$ times higher than DR2 with $\sim1\AA$ scatter for galaxies with $EW(H\beta)_{em} \geq -5\mbox{\AA}$ in DR2. We only examine galaxies with weak emission because they are more relevant to our study. The scatters are larger than the formal errors listed in Table~\ref{tab:sample} because the uncertainties in the calibration and modeling of spectra are not taken into account. Although a few galaxies may have fallen out of the selection criteria in DR3, they still possess a 'poststarburst'-like star formation history and will be shown to be so later in the paper.

\begin{table*}
\centering
	\begin{threeparttable}
		\caption{Poststarburst galaxy sample\label{tab:sample}}
		\begin{tabular}{cccccccc}
			\hline
			\hline
			ID & z & EW(H$\beta$)$_{em}$\tablenotemark{a,d} & \ewhd\tablenotemark{b} & \dn\tablenotemark{b} & EW(H$\beta$)$_{em}$\tablenotemark{a,d} & \ewhd\tablenotemark{b} & \dn\tablenotemark{b}\\
			&        & (\AA) & (\AA) &     & (\AA) & (\AA) & \\
			\hline
                & & & DR2 & & & DR3 & \\
                \hline
			133501  & 0.72916 & $<0.45$ &  5.30$\pm$0.12 & 1.57$\pm$0.01 & $<0.17$        & 3.57$\pm$0.33 & 1.54$\pm$0.02 \\
			144377  & 0.69650 & $<0.17$ & 4.25$\pm$0.15 & --\tablenotemark{c} & -0.43$\pm$0.05 & 4.42$\pm$0.17 & --\tablenotemark{c}\\
			209377  & 0.74703 & -1.41$\pm$0.13 & 4.67$\pm$0.09 & 1.45$\pm$0.01 & -1.96$\pm$0.07 & 5.34$\pm$0.10 & 1.39$\pm$0.01 \\
			210541  & 0.71477 & $<0.21$ & 7.02$\pm$0.17 & 1.39$\pm$0.01 & -1.53$\pm$0.07 & 7.07$\pm$0.15 & 1.32$\pm$0.01 \\
			260163  & 0.75285 & $<0.53$ & 7.51$\pm$0.26 & 1.34$\pm$0.01 & $<0.82$        & 8.26$\pm$0.29 & 1.35$\pm$0.02 \\
			\hline
		\end{tabular}
		\label{tab:psb}
        \tablenotetext{a}{Measured from continuum-subtracted spectra.}
        \tablenotetext{b}{Measured from emission-line-subtracted spectra.}
        \tablenotetext{c}{The spectra does not cover the wavelength range.}
        \tablenotetext{d}{Upper limits are 3$\sigma$ limits.}
	\end{threeparttable}

\end{table*}

Figure~\ref{fig:uvj} shows our sample on the UVJ diagram, a widely used technique to differentiate between quiescent and star-forming galaxies and identify poststarburst galaxies when only multi-wavelength broadband photometry is available. In this diagram, poststarburst galaxies are identified as quiescent galaxies with the bluest V-J colors \citep[e.g.,][]{whi12}, which are also likely to have high \ewhd. However, not all quiescent galaxies with high \ewhd\ have blue V-J colors. Using spectral features, we can detect poststarburst galaxies that may be missed by broadband photometry alone \citep{wu20}.

\subsection{Extracting kinematics of stars and ionized gas}

The deep LEGA-C spectra, taken with slits of north-south orientation and a pixel scale of 0\farcs2, offers exquisite detail with generally $<1"$ seeing \citep{str18}. For each row of the 2D spectra with median $S/N>2$ per pixel, we use pPXF to fit a model of the stellar continuum.

The stellar template is a non-negative, linear combination of \citep{vaz99} single stellar population (SSP) models, which are based on the Medium-resolution INT Library of Empirical Spectra \citep[MILES,][]{san06} with \citet{gir00} isochrones.

We mask out wavelength ranges where significant emission lines are present. Major lines are [\ion{O}{2}]$\lambda3727,3729$, H$\delta$, H$\gamma$, H$\beta$, and [\ion{O}{3}]$\lambda4959,5007$. The exact wavelength range is determined by visual inspection. It will be shown later in the paper that the line emission in each galaxy exhibits a variety of shapes and velocities. A common practice of modeling emission lines with Gaussian is not applicable to these galaxies. These fits yield spatially-resolved line-of-sight stellar velocity profiles along the north-south direction. Subtracting the best-fit stellar templates from the observed spectra gives us the line emission from the ionized gas along the slit.

\subsection{Deriving star-formation histories}

In this paper, we use the code BAGPIPES \citep{car18} to fit the observed LEGA-C spectra and broadband spectral energy distribution (SED) from UV to mid-IR to constrain the SFHs.

\subsubsection{Spectral-photometric fitting}
\label{sec:pipes}

\begin{table*}
	\centering
	\begin{threeparttable}
		\caption{Priors for the SFH fitting}
		\begin{tabular}{lllllcc}
			\hline
			Component & parameter & Symbol / Unit & range & prior & \multicolumn{2}{c}{Hyperparameters} \\
			\hline
			SFH & Stellar mass formed & $M_{\ast}/M_\odot$ & ($10^{10.5}$, $10^{12.5})$  & logarithmic & & \\
			& Stellar metallicity & $Z/Z_\odot$ & (0.1, 2.5) & logarithmic & &\\
			& decay timescale & $\tau$ / Gyr & (0.05, 4) & logarithmic & & \\
			& look-back time of formation & $t_{form}$ / Gyr & (4, $t_{Hubble}$) & uniform & & \\
			& look-back time of burst& $t_{burst}$ / Gyr & (0, 2) & uniform & & \\
			& burst fraction & $f_{burst}$ & (0, 1) & uniform & & \\
			\hline
			Dust & Dust attenuation at 5500\AA & $A_v$ / mag & (0, 8) & uniform & & \\
			& Power-law slope & n & (0.1, 2) & Gaussian & $\mu=0.7$ & $\sigma=0.5$ \\
			& Birth-cloud-to-diffuse attenuation & $\eta$ & (1, 5) & uniform & &\\
			\hline
			Calibration & 0th order & & (0.5, 1.5) & Guassian & $\mu=1$ & $\sigma=0.25$ \\
			& 1st order & & (-0.5, 0.5) & Guassian & $\mu=0$ & $\sigma=0.25$\\
			& 2nd order & & (-0.5, 0.5) & Guassian & $\mu=0$ & $\sigma=0.25$\\
			\hline
			Noise model & White noise scaling & & (0.1, 10) & logarithmic & & \\
			\hline
		\end{tabular}
		\label{tab:fit}

	\end{threeparttable}
\end{table*}

\begin{table*}
\hspace*{-1.5cm}
\begin{threeparttable}
	\caption{The star-formation histories of poststarburst galaxies}
\begin{tabular}{cccccccccc}
	\hline
	\hline
	ID & $\log(M_\ast/M_\odot)$ & $\tau$ & $t_{form}$& $t_{burst}$ & $f_{burst}$ & $A_v$ & $n$ & $\eta$ & $t_q$ \\
	   &                        &  (Gyr) &   (Gyr)    & (Gyr)       &            &    (mag)    &     &  & (Gyr)    \\
	\hline
	133501  & $10.99^{+0.03}_{-0.03}$ & $0.07^{+0.03}_{-0.02}$ & $6.45^{+0.40}_{-0.83}$ & $1.13^{+0.04}_{-0.04}$ & $0.42^{+0.09}_{-0.06}$ & $0.18^{+0.05}_{-0.04}$ & $0.86^{+0.23}_{-0.17}$ & $2.92^{+1.34}_{-1.23}$ & $0.57^{+0.03}_{-0.03}$ \\
	144377  & $11.07^{+0.02}_{-0.02}$ & $0.30^{+0.02}_{-0.02}$ & $5.57^{+1.08}_{-1.04}$ & $1.12^{+0.02}_{-0.02}$ & $0.98^{+0.01}_{-0.03}$ & $0.43^{+0.04}_{-0.05}$ & $1.03^{+0.07}_{-0.06}$ & $1.81^{+0.24}_{-0.19}$ & -- \\
	209377  & $11.14^{+0.02}_{-0.02}$ & $0.16^{+0.02}_{-0.02}$ & $4.74^{+1.01}_{-0.50}$ & $0.69^{+0.12}_{-0.10}$ & $0.15^{+0.08}_{-0.05}$ & $0.11^{+0.02}_{-0.02}$ & $1.31^{+0.15}_{-0.15}$ & $2.67^{+1.69}_{-1.40}$ & $0.23^{+0.04}_{-0.03}$ \\
	210541  & $11.25^{+0.04}_{-0.05}$ & $0.10^{+0.02}_{-0.01}$ & $4.56^{+0.80}_{-0.42}$ & $0.53^{+0.05}_{-0.05}$ & $0.23^{+0.03}_{-0.03}$ & $1.31^{+0.10}_{-0.11}$ & $0.50^{+0.07}_{-0.05}$ & $1.44^{+0.42}_{-0.30}$ & $0.05^{+0.01}_{-0.01}$ \\
	260163  & $10.94^{+0.04}_{-0.04}$ & $0.15^{+0.02}_{-0.03}$ & $5.35^{+1.00}_{-0.90}$ & $0.68^{+0.09}_{-0.09}$ & $0.48^{+0.17}_{-0.11}$ & $1.00^{+0.09}_{-0.08}$ & $1.15^{+0.08}_{-0.08}$ & $4.34^{+0.47}_{-0.67}$ & $0.15^{+0.02}_{-0.02}$ \\
	\hline
\end{tabular}
	\label{tab:sfh}
\end{threeparttable}
\end{table*}

\begin{figure*}
\centering
\includegraphics[width=0.95\textwidth]{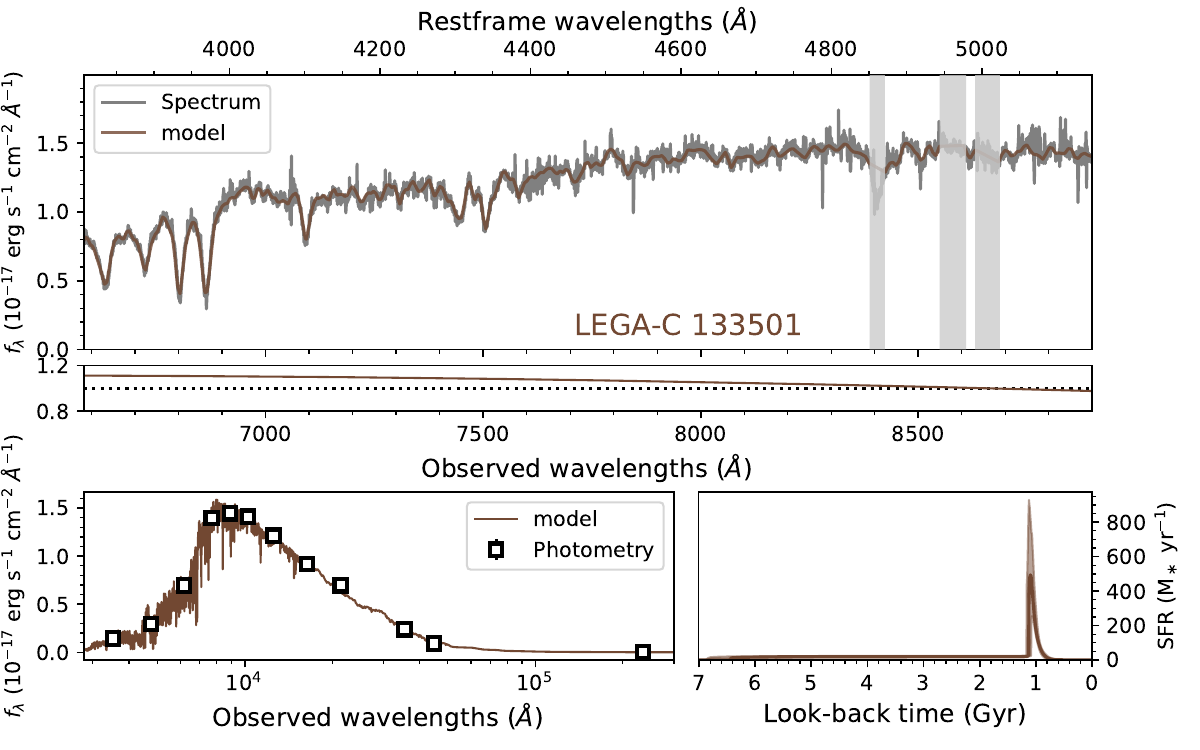}
\caption{The observed spectrum and photometry of LEGA-C~210541 and the best-fit model spectrum. Wavelength regions that may contain line emission (grey areas) are not fitted. The spectrum is multiplied by the factor shown in the middle panel to account for the mismatch to the broadband SED (see Section~\ref{sec:pipes}). The uncertainties of photometric measurements are smaller than the sizes of the squares in the bottom panel and thus not shown. The bottom right panel is the SFH posteriors. The line is the median and the shaded area shows the 16th and 84th percentiles of the posteriors. The fits of other galaxies are in the Appendix. \label{fig:fit}}
\end{figure*}
\begin{figure*}
\ContinuedFloat
\centering
\includegraphics[width=0.95\textwidth]{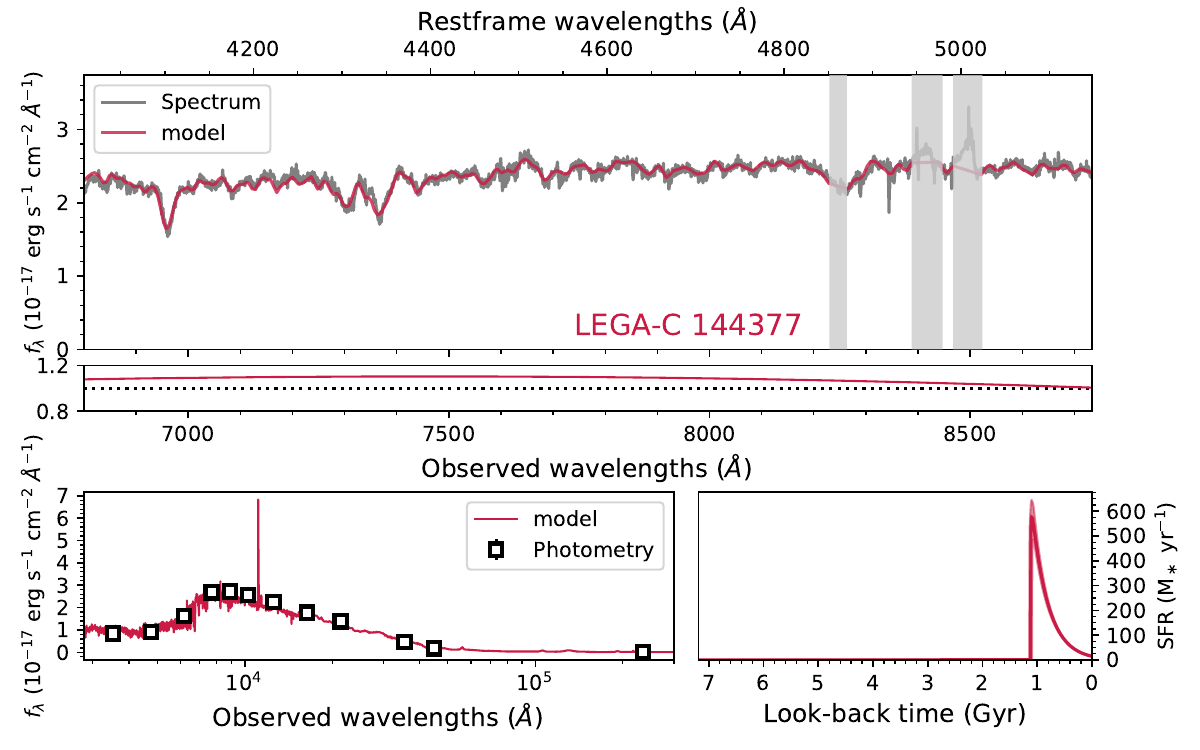}
\caption{Continue.}
\end{figure*}
\begin{figure*}
\ContinuedFloat
\centering
\includegraphics[width=0.95\textwidth]{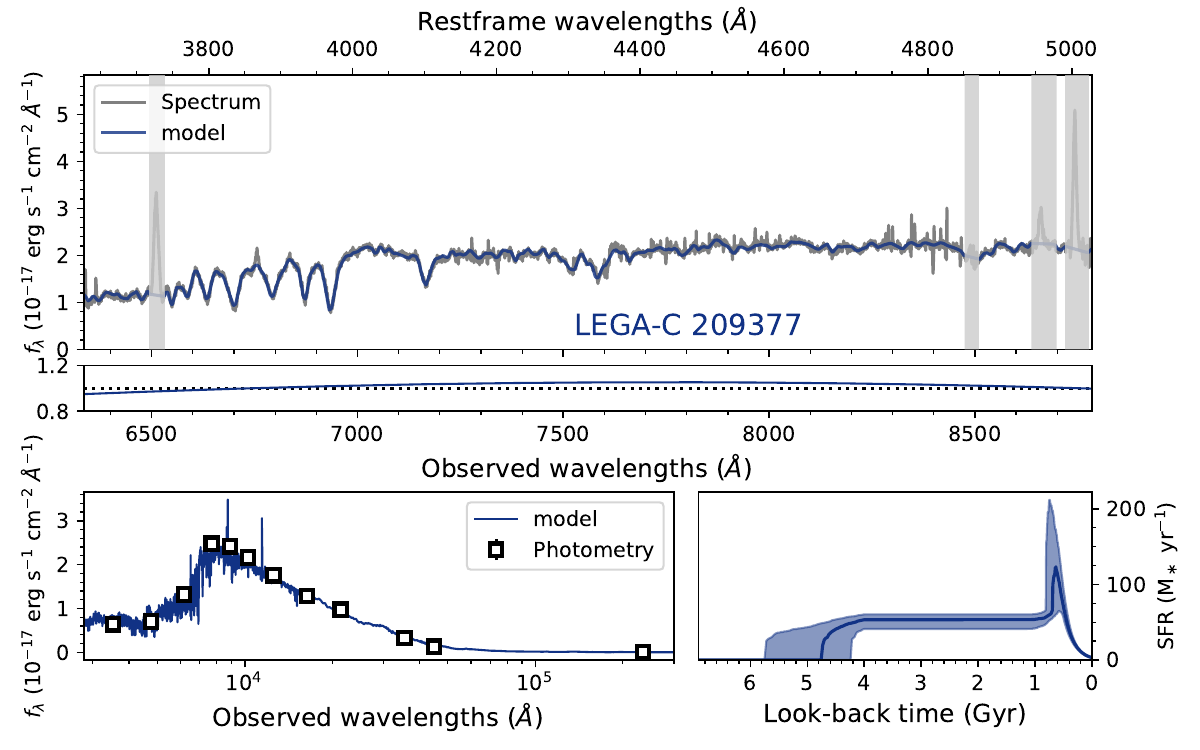}
\caption{Continue.}
\end{figure*}
\begin{figure*}
\ContinuedFloat
\centering
\includegraphics[width=0.95\textwidth]{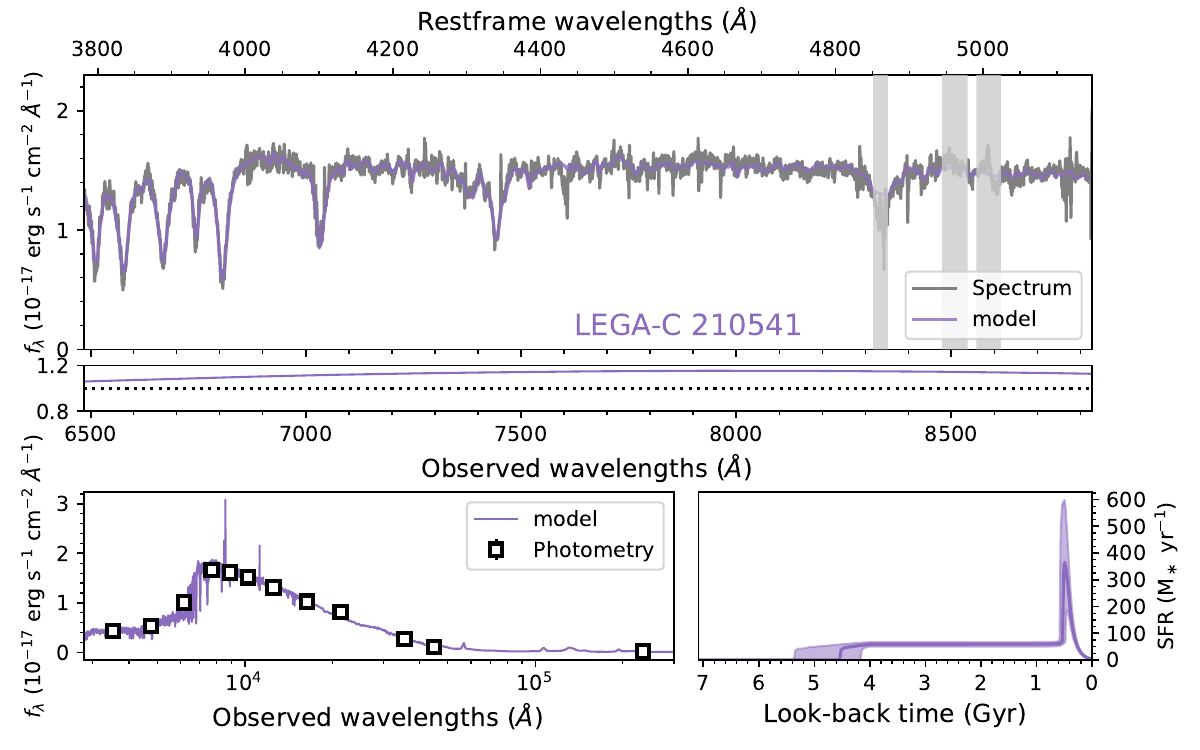}
\caption{Continue.}
\end{figure*}
\begin{figure*}
\ContinuedFloat
\centering
\includegraphics[width=0.95\textwidth]{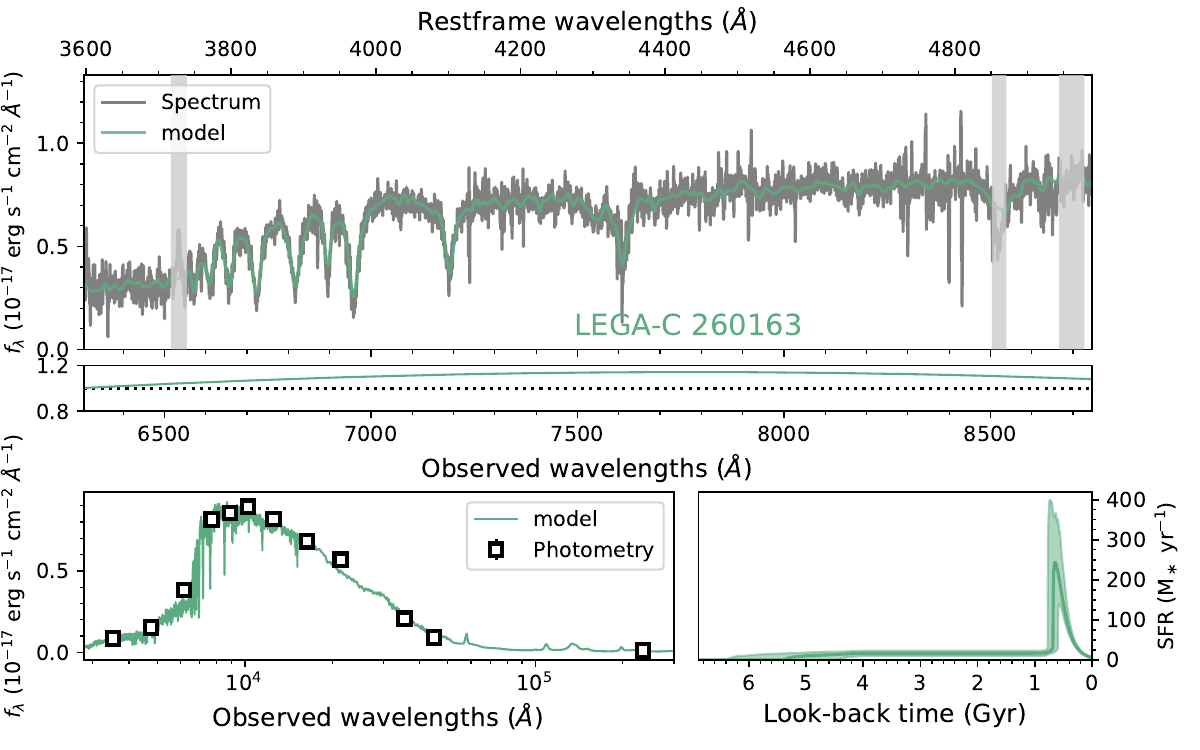}
\caption{Continue.}
\end{figure*}

\begin{figure}
\includegraphics[width=0.95\columnwidth]{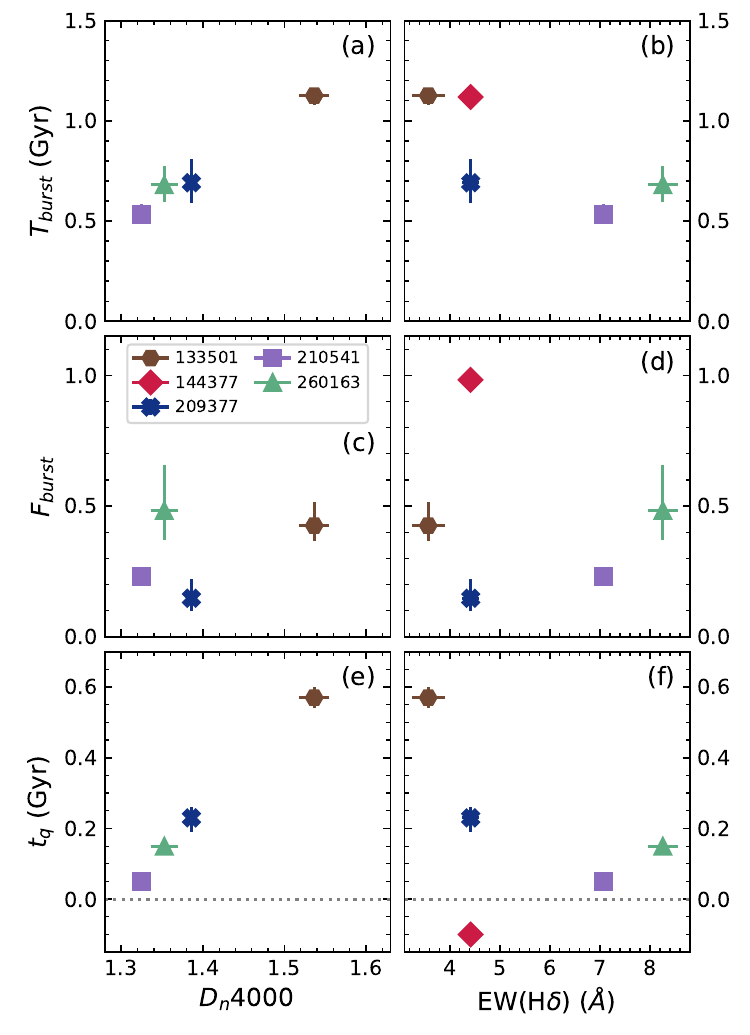}
\caption{The correlations between spectral indices and SFH parameters. The spectrum of LEGA-C~144377 does not cover the \dn\ region. \dn\ serves as a good qualitative proxy to the look-back time of the burst and the time that a galaxy became quiescent. On the other hand, there is no clear correlation between the burst strength and either \dn\ or \ewhd. LEGA-C~144377 has not become quiescent (see main text), we manually put it at $t_q = -0.1$. \label{fig:ind_sfh}   }
\end{figure}

Poststarburst galaxies, by the selection, are supposed to have rapidly falling recent SFRs. We assume that a poststarburst galaxy forms stars in two phases; a constant rate then followed by an exponentially-decay burst as described by the following parametric form:

\begin{equation} \label{eq:sfh}
SFR(t) \propto \left \{
\begin{array}{cr}
	\quad c,& t_{burst} \leq t \leq t_{form}\ ; \\
	e^{[-(t_{burst}-t)/\tau]}, & \ t < t_{burst} ,
\end{array}
\right.
\end{equation}
where $c$ is a constant, $t$ is the look-back time, $\tau$ is the timescale of the exponential decay, $t_{form}$ and $t_{burst}$ are the look-back times that the constant phase, and exponential phase starts, respectively. To ensure that the burst is represented by the exponential form rather than a short period of intensive constant star formation, we impose a minimum constant phase of 2~Gy, i.e., $t_{form} - t_{burst} \geq 2$~Gyr. Our tests indicate that extending the minimum length of the constant phase has no significant impact on our results. The relative SFRs in these two phases are controlled by the parameter $f_{burst}$:
\begin{equation}
	\frac{\int_0^{t_{burst}} e^{[-(t_{burst}-t)/\tau]} dt }{c \times (t_{form}-t_{burst})}= \frac{f_{burst}}{1 - f_{burst}}.
\end{equation}
$f_{burst}$ is the fractional stellar masses ever formed in the exponential phase thus between 0 and 1.

We employ the \citep{cf00} form of attenuation curve, characterized by $A_\lambda \propto \lambda^{-n}$, where the parameter $n$ is allowed to vary. We also allow the attenuation of stars in the birth cloud ($t<10^7$~yr) to be different from that of older stars with $A_{bc} = \eta \times A_{old}$. The dust emission models are taken from \citet{dra07}. The stellar continuum templates are generated with the updated \citet{bc03} stellar population synthesis model \citep{che16} using the MILES stellar spectral library \citep{fal11}. We assume all stars have the same metal content with a scaled-Solar abundance. The redshift and velocity dispersion values are fixed as in LEGA-C DR3 \citep{vdw21}.

The spectra are binned every 5 pixels before fitting using \textsf{SpectRes} \citep{car17}. The binning reduces processing time while preserving the constraining power because LEGA-C spectra have a pixel scale of $0.6/(1+z)\ \mbox{pixel}^{-1}$ at the restframe, and the MILES templates have a resolution of 2.5\AA. We then visually inspect the spectra and mask out wavelengths with line emission. We will show in Section~\ref{sec:ion_kin} that the emission lines in our poststarburst galaxies are highly asymmetric and at least partially driven by AGN, rendering the use of Gaussians implemented in BAGPIPES inadequate.

We include GALEX NUV, CFHT $u^*$, Subaru/HSC g, r, i, and z, VISTA/VIRCAM Y, J, H, and $K_s$, \textit{Spitzer}/IRAC ch1 and ch2 from the COSMOS2020 Classic catalog \citep{wea22} and Spitzer/MIPS 24$\mu$m from a matched IR source catalog of \citet{jin18} in our fitting\footnote{Part of the photometry is acquired by the COSMOS program, which can be accessed via \dataset[10.17909/T9XW2Q]{\doi{10.17909/T9XW2Q}}}. We set a floor uncertainty of 5\% for each photometric measurement. We add a second-order Chebyshev polynomial perturbation and a white noise model to the spectroscopic data to handle possible mismatch of the overall shape between the spectrum and the SED, as well as a possible underestimate of the uncertainties of the spectra. BAGPIPES takes a Bayesian approach to evaluate the most probable parameter sets. The priors are listed in Table~\ref{tab:fit}. Figure~\ref{fig:fit} shows an example of the fitting. We refer readers to \citet{car19b} for the details of the implementation.

\subsubsection{Star-formation histories of poststarburst galaxies}

The best-fit SFHs suggest that all poststarburst galaxies actually had a `burst', i.e., an increase of SFR in the past (Table~\ref{tab:sfh} and Figure~\ref{fig:ind_sfh}). In most cases, the starburst began 0.5-1~Gyr ago, produced 15-50\% of the stellar mass ever formed, and with exponentially decaying timescales of approximately 100~Myrs. Our poststarburst galaxies are at the very late stage of the starburst events, $> 4\tau$-decaying times after the burst. LEGA-C~133501 is even at $\sim10 \tau$-decaying times after the onset of the burst.

LEGA-C 144377 is a unique case among the sample galaxies. Essentially all its stellar masses formed in a prolonged burst with an e-folding time of 300~Myrs. To verify this result, various settings were modified in the fitting procedure, including the adoption of a different dust extinction law, exclusion of certain photometric bands or spectra in specific wavelength ranges, substitutions of the parametric SFH with a double power law, and use a non-parametric SFH. Despite these changes, all tests produced similar results, indicating that $\sim90\%$ of stars formed around 0.5-1~Gyr ago. We note that such a strong burst in poststarburst galaxies at similar redshifts has been presented before \citep{wil20,sue22}, indicating the diversity of poststarburst galaxies.

From the SFHs, we calculate $t_q$, the look-back time at which the star formation rate dropped below $10^{-10}$~yr$^{-1}$, a fiducial value to differentiate between star-forming and quiescent galaxies. The oldest poststarburst galaxies became quiescent less than 600 million years ago, while the majority have $t_q < 250$~Myr. Notably, LEGA-C~144377 is situated above this fiducial cut-off.

The correlations between SFH parameters and spectral indices are shown in Figure~\ref{fig:ind_sfh}. It is not surprising that \dn\ is strongly correlated with the onset of the burst and the time of quenching. However, neither \dn\ nor \ewhd\ is a reliable indicator of burst strength. Deriving this parameter requires spectral modeling.

We note that the assumed parametric form and priors do not enforce a higher peak SFR of the exponential component than the constant part. We have applied the poststarburst parametric SFH to fit a set of randomly selected quiescent galaxies with $\mbox{EW(H}\delta\mbox{)} < 2\AA$. The peak SFRs of the exponential components are either lower than the constant component or exceed the constant component for less than 10\% in the most prominent case, indicating a lack of a `burst' event in the recent past. The significant increase of SFRs of poststarburst galaxies in the fitting is necessary to match the observed data.

\subsection{Star-formation rates}

The SFRs is a crucial piece of information on poststarburst galaxies, nevertheless, hard to estimate. The peculiar SFHs question the applicability of conventional conversions from light to SFRs. The high incidence of AGN in poststarburst galaxies adds another source of uncertainties as AGN could contribute to almost all wavelengths used for estimating the SFRs. Here we calculate 4 SFRs based on data at different wavelengths which trace various physical processes, following conventional calibrations and methodologies. Considering the potential contribution from AGN, the values reported here should be considered as upper limits.

Firstly, we derive the SFRs by averaging the reconstructed SFHs over the last 100~Myrs. With a timescale of $\sim100$Myrs for the $\tau$-decay (Table~\ref{tab:sfh}), the instantaneous SFRs are roughly half of the listed values.  Next, we calculate the SFRs based on dust-corrected H$\beta$ emission, using the calibration by \citet{ken12} and the \citet{cha03} IMF, with an assumption of an intrinsic flux ratio of $H\alpha/H\beta = 2.86$. We use the best-fit $A_V$, $n$, and $\eta$ to calculate the attenuation of stars with $<10$~Myr at the wavelength of H$\beta$ and use the value as the attenuation of the H$\beta$ emission line.

Additionally, we determine the SFRs from the restframe 2800~\AA\ UV and 8-1000~$\mu$m IR luminosities of the best-fit SED models, using the prescription by \citet{bel05} and a \citet{cha03} IMF. The SFRs are computed as $\Psi[M_\odot\ \mbox{yr}^{-1}] = 1.09\times10^{-10} (L_{IR} + 2.2 L_{UV}) [L_\odot]$. We will demonstrate in Sec.~\ref{sec:sfr} that the old stars in poststarburst galaxies can contribute a large fraction of energy that heats the dust. This common calibration can be an overestimate.

Unlike the UV luminosity, which is relatively well-constrained by the NUV and $u^*$ bands, the total IR luminosities are only sampled by a single \textit{Spitzer} MIPS 24$\mu$m measurement. To verify the result, we identify LEGA-C~201541 and LEGA-C~260163 in the far-IR source catalog of \citet{jin18} that are detected at 100~$\mu$m with an S/N$>4$. We re-fit with the addition of 100~$\mu$m photometry and find that the SFRs change by less than a factor of 2, which does not affect our conclusions.

Finally, we cross-match the poststarburst galaxies with the COSMOS VLA 3~GHz source catalog \citep{smo17}, with a maximum matching radius of 0\farcs7 \citep{bar17}. We convert the radio continuum fluxes to SFRs using the \citet{bel03} calibration of the radio-FIR correlation and report the values in Table~\ref{tab:sfr}. We note that the contribution of AGN to the 3~GHz luminosity cannot be ignored. In particular, the inferred $SFR_{3GHz}$ for LEGA-C~209377, which exceeds $2000\ \mbox{M}_\odot\ \mbox{yr}^{-1}$, is almost for sure contaminated by AGN. This galaxy has a 3~GHz luminosity of $8.6\times10^{24}\ \mbox{W}\ \mbox{Hz}^{-1}$ \citep{bar17}, which is significantly higher than the commonly-used threshold for identifying radio-loud AGN \citep[$10^{23}\ \mbox{W}\ \mbox{Hz}^{-1}$,][]{bes05}. The 3~GHz luminosities of all other galaxies in our sample are all lower than $10^{23}\ \mbox{W}\ \mbox{Hz}^{-1}$.

Table~\ref{tab:sfr} reveals an intriguing disagreement among the SFRs derived by different methods. Notably, the SFRs estimated using the UV+IR and 3~GHz continuum emission appear to be consistently higher, potentially resulting in the misclassification of some poststarburst galaxies as part of the star-forming main sequence. We will discuss this discrepancy in Section\ref{sec:sfr}.

\begin{table}
\caption{Star Formation Rates}
\centering

		\begin{tabular}{ccccc}
			\hline
			\hline
			ID & SED & $H\beta$ & UV+IR & 3~GHz \\
			\hline
			133501  & $0.0^{+0.1}_{-0.0}$ & (0.2) & $2.9^{+0.7}_{-0.6}$ & $17\pm3$  \\
			144377  & $17.8^{+1.8}_{-1.8}$ & $1.0\pm0.1$ & $24^{+2}_{-3}$ & $31\pm3$ \\
			209377  & $4.4^{+0.9}_{-0.6}$ & $2.7\pm0.1$ & $7.7^{+0.9}_{-0.7}$  & $2733\pm142$  \\
			210541  & $6.0^{+2.3}_{-1.5}$ & $3.8\pm0.2$  & $40^{+4}_{-5}$  & $42\pm3$  \\
			260163  & $8.0^{+1.9}_{-2.0}$ & (12.1) & $19^{+3}_{-2}$  & (27) \\
			\hline
		\end{tabular}
		\label{tab:sfr}
		\begin{tablenotes}
			\item SFRs in $M_\odot$~yr$^{-1}$. Values for the `SED' and `UV+IR' are based on the 16th, 50th, and 84th percentiles of the spectrophotometric fitting. Values for `H$\beta$' and `3~GHz' are derived from the fluxes and their uncertainties, respectively. Here we do not attempt to quantify the fluxes contributed by AGN. Numbers in parentheses are 3-$\sigma$ upper limits.
		\end{tablenotes}

\end{table}

\subsection{ALMA observations}
\label{sec:alma}

\begin{table*}
\caption{ALMA observation}
	\hspace*{-2cm}
	\scriptsize
		\begin{tabular}{ccccccccc}
			\hline
			\hline
			ID & z & $T_{int}$ & beam size & beam P.A. & $S_{CO(2-1)\Delta v}$ & $L_{CO}$ & $\log(M_{H2}/M_\odot)$\tablenotemark{a} & $M_{H2}/M_\ast$\\
			 &   &  (min)    &     &   ($\degree$)   &      (Jy km~s$^{-1}$)    & ($10^9$ K km s$^{-1}$ pc$^2$)    &   &   \\
			\hline
			133501  & 0.72916 & 146 & $2\farcs2\times1\farcs7$ & 83    & $<0.125$        & $<0.89$       & $<$9.68 & $<0.05$  \\
			144377  & 0.69650 & 106 & $2\farcs2\times1\farcs5$ & $-87$ & 0.414$\pm$0.040 & 2.68$\pm$0.26 & 10.17$\pm$0.04 & 0.13 \\
			209377  & 0.74703 & 126 & $1\farcs4\times1\farcs2$ & 76    & 0.077$\pm$0.021 & 0.57$\pm$0.16 & 9.50$\pm$0.12 & 0.02  \\
			210541  & 0.71477 & 86  & $2\farcs2\times1\farcs8$ & 75    & 0.375$\pm$0.026 & 2.56$\pm$0.18 & 10.15$\pm$0.03 & 0.08 \\
			260163  & 0.75285 & 115 & $1\farcs4\times1\farcs2$ & 71    & 0.491$\pm$0.056 & 3.72$\pm$0.42 & 10.31$\pm$0.05 & 0.23  \\
			\hline
		\end{tabular}
		\label{tab:alma}

		\tablenotetext{a}{Assuming $r_{21}=0.8$ and $\alpha_{CO}=4.0$ }

\end{table*}

\begin{figure}
	\includegraphics[width=0.95\columnwidth]{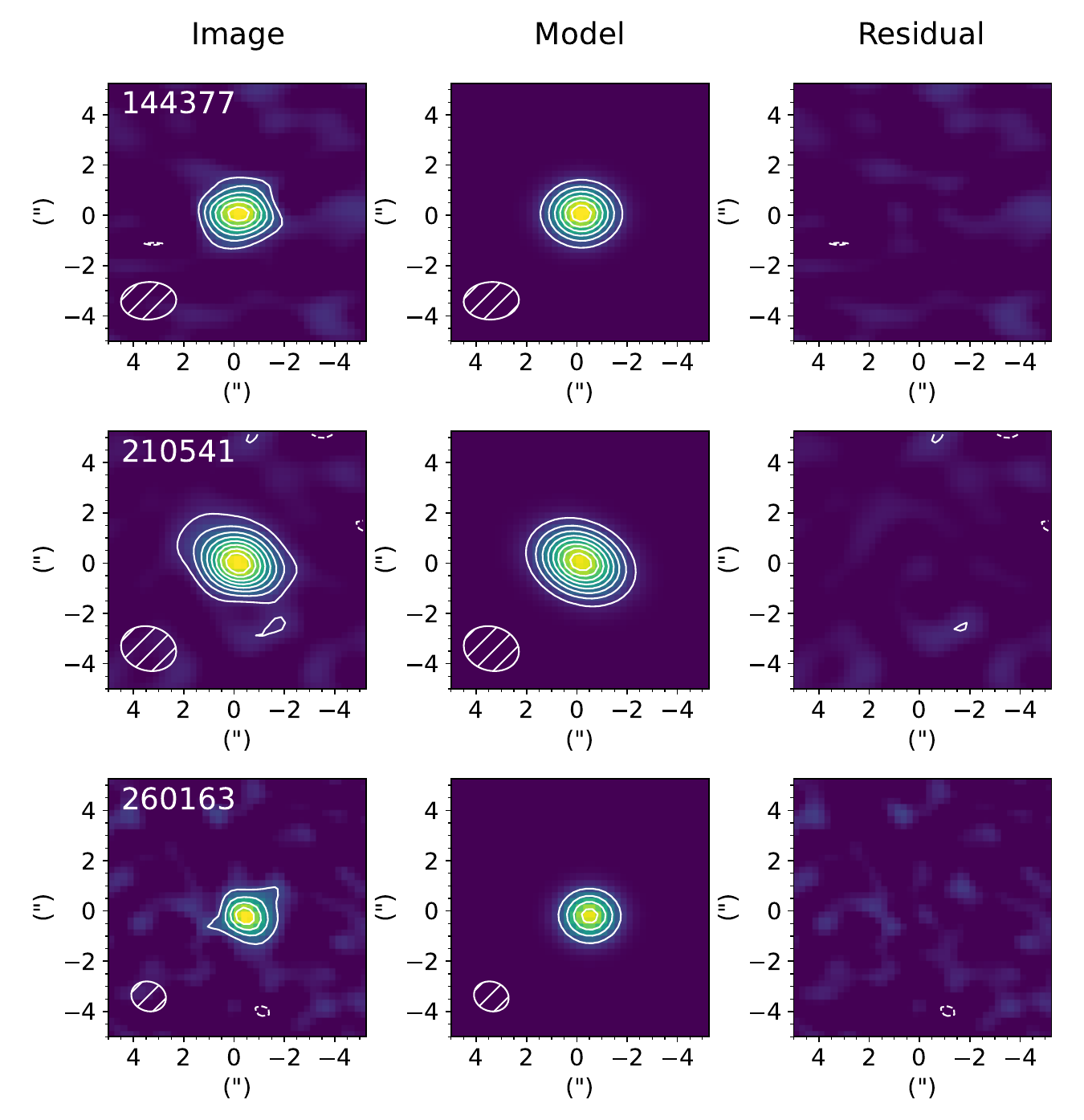}
	\caption{The moment 0 maps of CO(2-1) and best-fit Gaussian models. Contours increase every 3$\sigma$, starting from 3$\sigma$, where $\sigma$ is measured from the central 3” of the residual maps. The beam sizes are plotted at the lower left corners. All 3 CO(2-1) emissions can be modeled by a single Gaussian component, with no clear sign of disturbed morphology or outflow under the current resolution.\label{fig:mom0} }
\end{figure}

The ALMA observations were carried out in program 2019.1.00702.S (PI: P. Wu) in January and February 2019 using the ALMA Band 4 receivers \citep{asa14}. The total on-source integration time for each target is 1.5-2.5 hours. The data are reduced and imaged using the standard ALMA pipeline. The spatial resolutions range from 1\farcs2-2\farcs2 (9-16 kpc) with a robust parameter of 0.5 (Table~\ref{tab:alma}).

The CO(2-1) emission is clearly detected in three galaxies, LEGA-C 144377, 210541, and 260163, around their optical centers at an 8-15$\sigma$ level. Our visual inspection suggests that the CO emission is marginally resolved and we model the emission as a 2D elliptical Gaussian using the CASA task \texttt{IMFIT}, which measures the apparent sizes and then performs deconvolution to constrain the intrinsic sizes of sources. For LEGA-C~144377 \texttt{IMFIT} fails to obtain an intrinsic size because the measured image size is similar to the beam size. The other two galaxies both have major axes of $\sim1"$ after deconvolution (Table~\ref{tab:size}). All three galaxies are well represented by the Gaussian model (Figure~\ref{fig:mom0}). We also fit the sources either assuming a disk model or in the visibility space,  the results are consistent within the uncertainties.

\begin{deluxetable*}{|cc|ccc|ccc|ccc|}
\tabletypesize{\scriptsize}
\tablecaption{Deconvolved sizes of ALMA sources}
\tablehead{
\colhead{} & \colhead{} & \colhead{} & \colhead{144377} & \colhead{} & \colhead{} & \colhead{210541} & \colhead{} & \colhead{} & \colhead{260163} & \colhead{}
}
\startdata
    &         & Flux\tablenotemark{a}    & Size\tablenotemark{b} & P.A.          & Flux\tablenotemark{a}    & Size\tablenotemark{b} & P.A.        & Flux\tablenotemark{a}    & Size\tablenotemark{b} & P.A.       \\
    &         &         &  (") & ($^{\circ}$)  &         &  (") & ($^{\circ}$)&         &  (") & ($^{\circ}$)  \\
\hline
ALMA CO(2-1) &  beam   &  & $2.20\times1.49$ & -87 & & $2.23\times1.75$ & 75 & & $1.38\times1.16$ & 71 \\
robust=0.5     &   source & 420(39) & - & - & 375(26) & $1.26(0.13)\times0.60(0.42)$ & 57(26) & 494(57) & $1.06(0.30)\times0.85(0.33)$ & 90(74)\\
\hline
ALMA CO(2-1)&  beam   &  & $2.60\times2.04$ & -88 & & $2.66\times2.18$ & 74 & & $1.80\times1.50$ & 61 \\
Natural     &   source & 405(31) & - & - & 386(17) & $1.38(0.18)\times0.92(0.18)$ & 64(22) & 576(48) & $1.70(0.25)\times0.92(0.18)$ & 81(36)\\
\hline
ALMA CO(2-1)&  beam   &  & $1.64\times0.83$ & -88 & & $1.49\times1.23$ & 70 & & $1.13\times0.93$ & -89 \\
Uniform &   source & 350(93) & - & - & 279(28) & - & - & 402(76) & - & - \\
\hline
VLA 3~GHz &       beam   &         & $0.75\times0.75$ & &       &  $0.75\times0.75$    &      &         &  $0.75\times0.75$    &    \\
        &   source & 32.7(4.9) & $0.59(0.17)\times0.07(0.25)$ & 29(14) & 42.9(4.1) & - & - & 0.7(5.9) & - & - 
\enddata
\tablecomments{Numbers in the parentheses are uncertainties.}
\tablenotetext{a}{The unit for VLA 3~GHz flux density is $\mu$Jy. The unit for ALMA CO(2-1) flux is mJy~km~s$^{-1}$. }
\tablenotetext{b}{The sources sizes and P.A. are deconvolved sizes derived with CASA task \texttt{IMFIT}. Sources that are smaller than the beam have no deconvolved values. }
\label{tab:size}
\end{deluxetable*}

To further examine the robustness of the size measurements, we make the images using both natural weighting and uniform weighting and repeat the modeling with elliptical Gaussian (Table~\ref{tab:size}). With uniform weighting, which yields better angular resolutions but lower signal-to-noise ratios, LEGA-C~210541 and 260163 have $\sim20\%$ lower model fluxes than those measured with \texttt{robust=0.5} and become apparently unresolved despite the better angular resolutions. We have examined the residual images to confirm that the Gaussian models represent the sources well, as done in Figure~\ref{fig:mom0}. These results suggest the existence of faint and extended emission that is missed in the uniform-weighted maps with lower signal-to-noise ratios. In addition, with the more sensitive but lower angular resolution images with natural weighting, the flux and intrinsic size of LEGA-C~260163 become marginally larger, again implying faint and diffuse emission.

To compromise between sensitivity and angular resolution, we choose to use the \texttt{robust=0.5} map for this paper. We then measure the fluxes by integrating the data cubes over a 3" radius from the optical centers, then used either one or two Gaussian components to fit the integrated spectra. We used the best-fit Gaussian models and their uncertainties as the fluxes of the CO(2-1) emission, as reported in Table~\ref{tab:alma}. The results are in agreement with modeling the images (Table~\ref{tab:size}). Additionally, we calculated the direct sum over the velocity range where the line emission existed based on visual inspection, and the fluxes from this method are consistent with the Gaussian fits.

The fourth galaxy, LEGA-C~209377, exhibits weak CO(2-1) emission 1" northwest of the optical center. We follow a similar method for measuring its flux, integrating over a 2" x 2" region centered at the peak of the emission, and fitting a Gaussian to the spectrum. LEGA-C~133501 had no detectable CO(2-1) emission. We estimated the upper limit of the flux using a single $\pm400$\kms\ channel.

To calculate H$_2$  masses, we assumed $r_{21}=0.8$ and a Milky-Way-like CO–H$_2$  conversion factor of $\alpha_{CO} = 4.0$, as per \citet{bol13}. The molecular gas masses range from $2\times10^9 M_\odot$ to $3\times10^{10} M_\odot$. The one undetected source, LEGA-C~133501, has a 3$\sigma$ upper limit of $4.8\times10^9 M_\odot$. The ratio between the masses of molecular gas and stars, $f_{H2} \equiv M_{H2}/M_\ast$, range between 2\% to 26\% and are comparable to the values found in previous studies \citep{sue17,bez22,zan23}

\section{Spatial distribution of stars, molecular gas, and star-formation}
\label{sec:dist}
\begin{figure*}
	\includegraphics[width=0.95\textwidth]{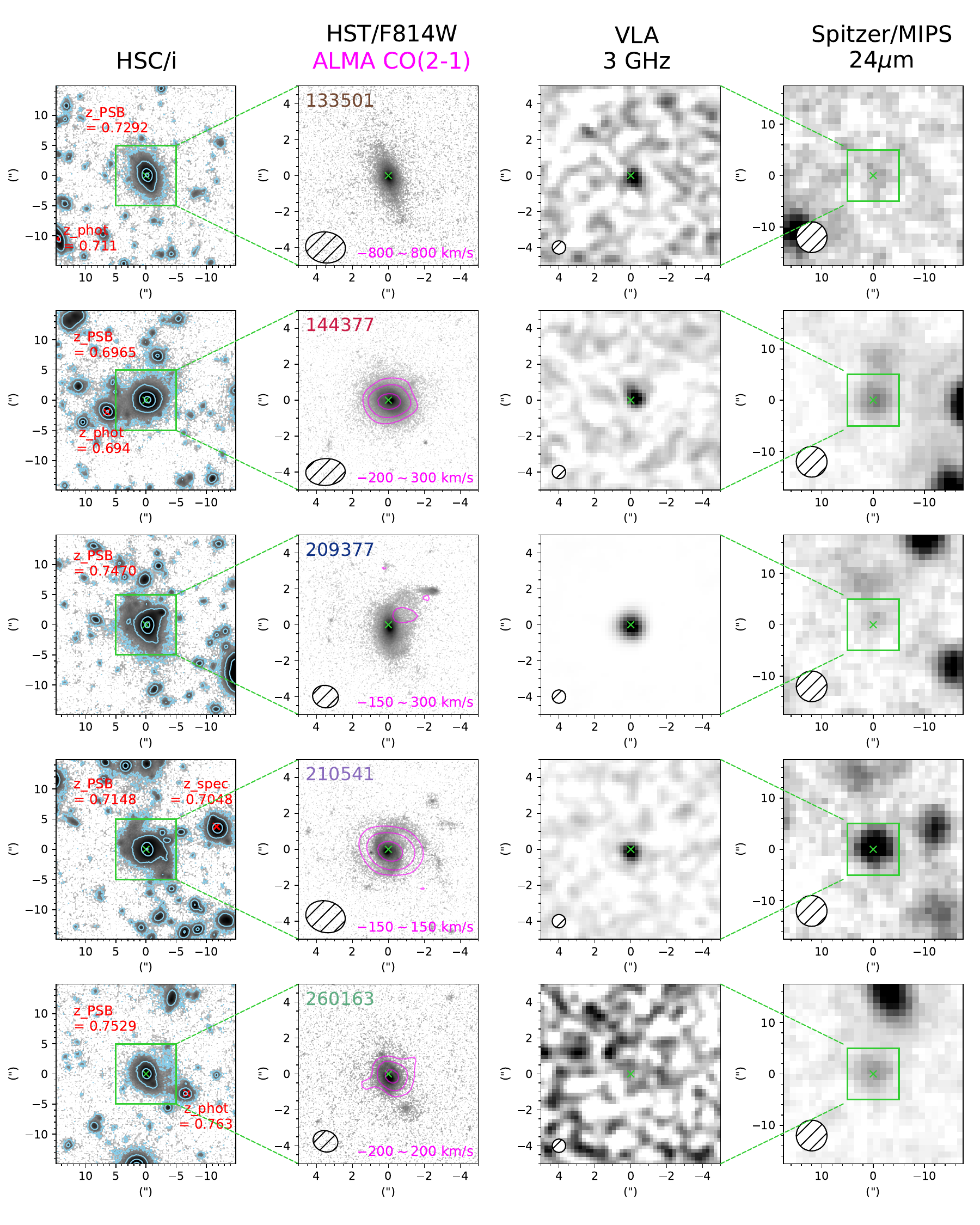}
	\caption{The distributions of star, gas, and star formation of poststarburst galaxies. \textit{First column:} HSC $i$-band light. The contour levels are 31, 29, and 27 mag/". Galaxies with similar redshifts to poststarburst galaxies are labeled. Most poststarburst galaxies have clear tidal features or a companion in the field of view. \textit{Second column:} \textit{HST}/F814W (gray scale image) and CO(2-1) (contours) maps. The morphologies of the main stellar bodies of poststarburst galaxies are vastly different. Contour levels are -3, 3, 6, 12, and 18 $\sigma$. Hashed ellipses are the primary beams. \textit{Third column:} The VLA 3~GHz maps. Radio emissions are unresolved. \textit{Fourth column:} \textit{Spitzer} MIPS 24$\mu$m maps. Galaxies with CO(2-1) detections are also detect in 24$\mu$m. Hashed circles in the 3rd and 4th columns are the nominal angular resolutions.}
	\label{fig:cutout}
\end{figure*}

Figure~\ref{fig:cutout} presents an overview of the spatial distribution of stars, gas, and star formation in post-starburst galaxies through multiwavelength images including Subaru HyperSuprimCam (HSC) $i$, \textit{HST}/F814W, ALMA CO(2-1), VLA 3~GHz, and \textit{Spitzer}/MIPS 24$\mu$m maps of our sample. This figure provides a comprehensive look at the distribution of these different components across poststarburst galaxies and can be used to gain insights into the processes that led to the cessation of star formation in these galaxies.

\subsection{Stars}
High-resolution images from the \textit{HST} reveal a diverse array of morphologies among poststarburst galaxies, including disk, elliptical, and tidal features. 
LEGA-C~210541, for instance, is oriented face-on and displays two distinct tidal arms. LEGA-C~209377 presents a low surface brightness fan-shaped structure connecting a fainter stellar body located approximately 3\arcsec northwest. LEGA-C~260163 has a small companion galaxy located around 3\arcsec southwest. These 3 galaxies are those showing more recent bursts. Conversely, the \textit{HST} images of LEGA-C 133501 and 144377 do not exhibit any clear tidal features. These galaxies have older bursts that occurred more than 1~Gyr ago.

Deeper images from the HSC reveal even more low surface brightness features that are not visible in the \textit{HST} images, extending as far as 40~kpc from the galaxy centers. For example, the two prominent tidal arms of LEGA-C~210541 are only visible out to about $\sim2\arcsec$ from the galaxy center in the \textit{HST} image but continue to about $\sim5\arcsec$ (or approximately 37~kpc) in the HSC image. Similarly, LEGA-C~209377 shows a fan-shaped feature at its northwest in the \textit{HST} images, but the HSC image reveals a wider, lower surface brightness, and asymmetric light distribution toward the northeast. LEGA-C~144377 appears to be a normal spheroid galaxy in the \textit{HST} image, but the HSC image again reveals an asymmetric low-surface-brightness feature at the southeast of the galaxy.

When viewed from a wider field of view, it is common to find nearby companions. Four of the poststarburst galaxies have close neighbors at a similar cosmological distance, estimated from either photometric or spectroscopic redshifts. The $K_s$ band flux ratios range from 1:2 to 1:5 and the separations are between 50~kpc and 130~kpc. We investigate 1251 galaxies with similar stellar masses ($M_\ast > 10^{11} M_\odot$) and photometric redshifts ($0.6 < z_{phot} < 0.8$) in the COSMOS2020 catalog and find that only 199 (or 16\%) have a companion within 130 kpc, with a $\Delta z_{phot} < 0.02$, and a $K_s$ flux ratio exceeding 1/5. The probability that our poststarburst galaxy was randomly selected from the larger sample is just a mere 0.3\%. This suggests that the interactions with nearby galaxies may have played a role in the formation of poststarburst galaxies, although a larger sample with dense spectroscopic redshifts will be needed to further verify this result.

\subsection{Star formation and molecular gas}
\label{sec:sfgas}
The third and fourth columns of Figure~\ref{fig:cutout} are two commonly used tracers of star formation: VLA 3~GHz and \textit{Spitzer}/\textit{MIPS}~24$\mu$m. The \textit{Spitzer}/\textit{MIPS}24$\mu$m has a low resolution of $\sim6\arcsec$. Three galaxies (LEGA-C~210541, 144377, and 260163) have clear 24$\mu$m sources that coincide with the optical centers. LEGA-C~209377 has a weak ($\sim4.5\sigma$) detection at $\sim1\arcsec$ northwest of the optical center, which also matches the weak CO(2-1) detection.

Four poststarburst galaxies have 3~GHz emission at their optical centers. LEGA-C~209377 is clearly a radio-loud AGN given its high luminosity. For other sources, we model their emission as 2D elliptical Gaussian and constrain their intrinsic sizes using \texttt{IMFIT}, assuming a circular beam of $0\farcs75\times0\farcs75$ \citep{smo17}. LEGA-C~144377 has an elongated shape with an intrinsic major axis of $\sim0\farcs6$ and the other two sources (LEGA-C 133501 and 210541) are point-source-like If the radio emission is powered by star formation activities, this compactness implies that the star formation concentrates within a small region, rather than being spread out over a large area. Sub-kpc scale residual star formation has been shown to be prevalent at the center of local poststarburst galaxies \citet{wu21a}.



In addition to 3~GHz and 24$\mu$m emission, later in Section~\ref{sec:ion_kin}, we will show from the H$\beta$ emission that LEGA-C~210541 has a central star-forming region with an apparent angular size of $\sim0\farcs8$, which is comparable to the seeing of our spectroscopic observation. Therefore, the 3 different tracers all suggest that the star formation activities in poststarburst galaxies, if exist, concentrate at galaxy centers.

Closely related to star formation is the raw material that fuels it: molecular gas. The distributions of the CO(2-1) emission are plotted in contours in Figure~\ref{fig:cutout}. The velocity ranges used to integrate the CO(2-1) emission are listed in each panel.

LEGA-C~209377 exhibits a weak CO(2-1) detection of $3.5\sigma$ significance, located at a distance of $\sim1\arcsec$ northwest of its optical center. Interestingly, this detection coincides with a fan-shaped stellar stream connecting a small companion galaxy located $\sim3\arcsec$ away. Additionally, there is a tentative source of $\sim4\sigma$ significance at 24$\mu$m at the same location. These observations suggest that this galaxy has recently experienced gravitational disturbance, which may have triggered off-center star formation and possibly led to quenching.

The CO(2-1) emission in LEGA-C 210541, 144377, and 260163 is well-detected around the optical centers. LEGA-C~210541 and LEGA-C~260163 have intrinsic sizes of $1\farcs1$ ($\sim8$~kpc) and $1\farcs3$ ($\sim10$~kpc), respective. The CO emission observed in these 2 galaxies exhibits a spatial distribution that extends beyond the confines of the spatially-unresolved star-formation activities. This observation suggests that the molecular gas present in these galaxies may not be linked to star formation.

The centers of the CO(2-1) emission in LEGA-C~144377 and LEGA-C~210541 align closely with their respective optical centers, with an offset of approximately $\sim0\farcs2$. Conversely, the center of the CO emission in LEGA-C~260163 is located $0\farcs36\pm0\farcs08$ ($\sim2.7\pm0.6$~kpc) southwest of its optical center. Given the reasonable detection ($S/N \simeq 9$), the offset is small compared to the synthesized beam size of 1\farcs4 and yet still significant. This offset suggests that the molecular gas is recently disturbed and not yet dynamically settled.

\section{Kinematics}
\label{sec:kin}
With the ALMA observation and the ultra-deep optical slit spectra, these galaxies provide us a rare opportunity to look into the kinematics of stars, ionized gas, and molecular gas simultaneously.

\subsection{Kinematics of ionized gas}
\label{sec:ion_kin}

\begin{figure*}
	\includegraphics[width=0.95\textwidth]{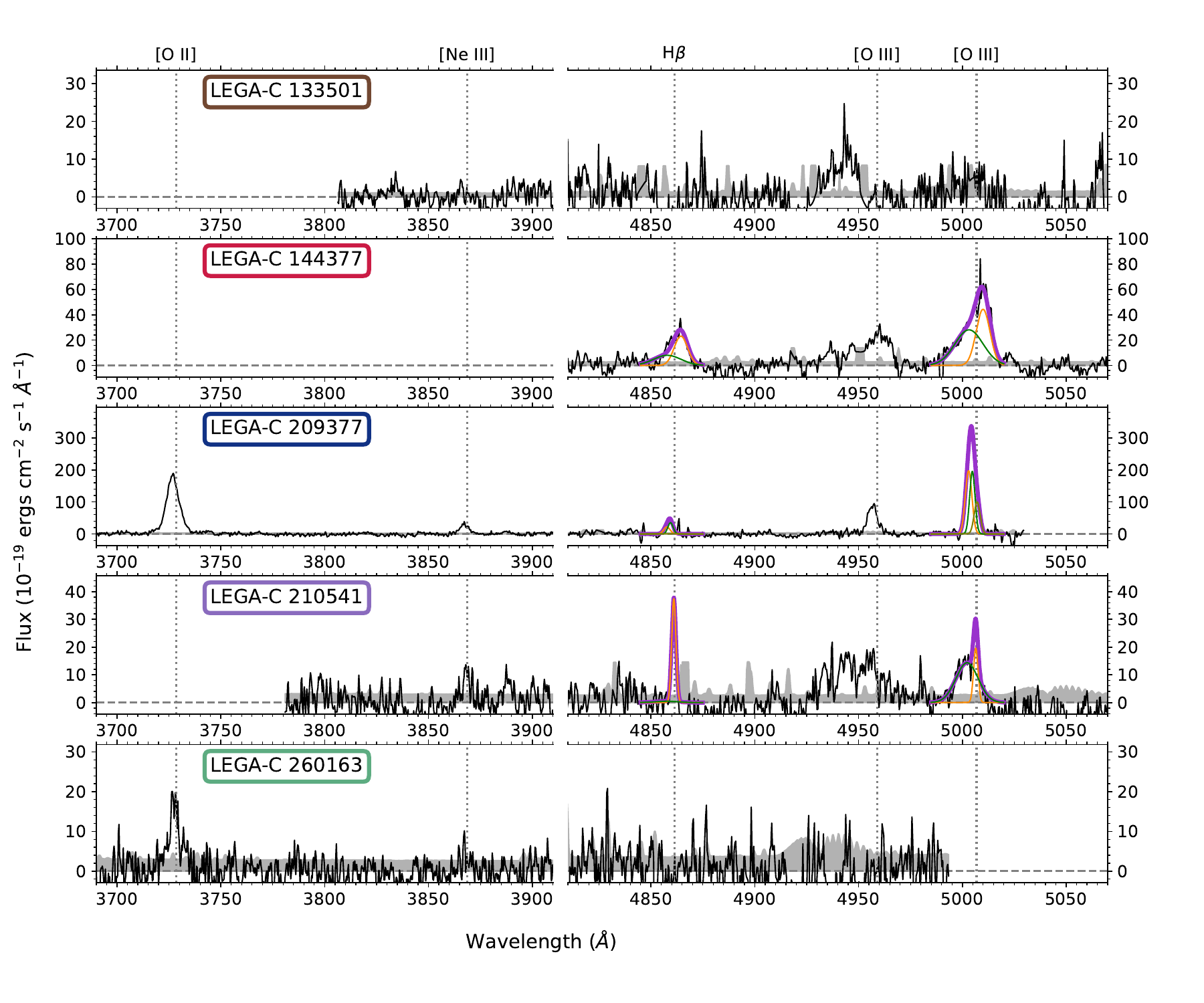}
	\caption{The continuum-subtracted spectra of poststarburst galaxies. The thick magenta lines are the best-fit multiple Gaussian models to the H$\beta$ and \othreea\ lines. Thin orange, green, and gray lines are individual Gaussian components. The \othreea\ lines show asymmetric broad components.}
	\label{fig:em}
\end{figure*}

\begin{table*}
	\begin{threeparttable}
		\caption{Properties of emission lines}
		\begin{tabular}{cccccccc}
			\hline
			\hline
LEGA-C ID & component & $f_{O3}$ & $f_{H\beta}$ & $f_{O2}$& Velocity & FWHM & $\log(O3/Hb)$ \\
\hline
       &   & \multicolumn{3}{c}{($10^{-19}$ erg s$^{-1}$ cm$^{-2}$)} & \multicolumn{2}{c}{(km s$^{-1}$)} &  \\
\hline
144377 & A & $370\pm36$ & $187\pm11$ & — & $186\pm7$ &  $469\pm26$ & 0.42 \\
       & B & $477\pm35$ & $132\pm16$ & — &$-218\pm54$ & $955\pm63$ & 0.68 \\
       & total & $823\pm12$ & $303\pm9$ & — & — & — & 0.55 \\
209377 & A & $777\pm416$ & $84\pm54$ & — &$-230\pm51$ & $221\pm34$ & 1.09\\
       & B & $633\pm55$ & $111\pm56$ & — &$-123\pm15$ & $182\pm39$ & 0.88\\
       & C & $359\pm135$ & $6\pm25$ & — &$15\pm37$ & $202\pm28$ & 1.90 \\
       & total & $1787\pm13$ & $212\pm9$ & $1412\pm8$ & — & — & 1.07 \\
210541 & A & $53\pm6$ & $97\pm4$ & — &$-24\pm3$ & $151\pm8$ & -0.14 \\
       & B & $193\pm11$ & $7\pm9$ & — &$-275\pm31$ & $755\pm77$ & 1.59 \\
       & total & $228\pm10$ & $96\pm5$ & — & — & — & 0.50 \\
260163 & total & — & — & $124\pm8$ & — & — & — \\
\hline
		\end{tabular}
		\begin{tablenotes}
			\item The flux of each component is derived from multiple Gaussian fits. Total fluxes are calculated from summing the spectra over the wavelength range used for fitting (Figure~\ref{fig:em}) for \othreea\ and H$\beta$ and $\pm800$ \kms\ for [\ion{O}{2}].
		\end{tablenotes}
		\label{tab:em}
	\end{threeparttable}
\end{table*}

 \begin{figure}
	\includegraphics[width=0.95\columnwidth]{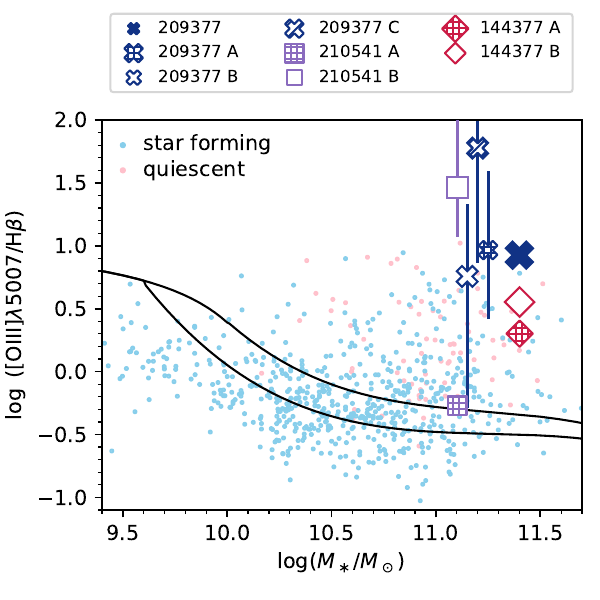}
	\caption{The \othreea/H$\beta$ flux ratios and stellar masses of each emission component of poststarburst galaxies. For LEGA-C~209377, the sum of components are also plotted. The shapes of markers represent different galaxies and the open, filled, and hashed symbols are different kinematic components in a galaxy. The stellar masses of LEGA-C~209377 are shifted by hand for the presentation. Black lines separate galaxies dominated by star-forming regions (bottom) from AGN (top), as well as the mixed region in between \citep{jun14}. Small dots are other LEGA-C galaxies where the flux ratios can be measured. Blue and red represent star-forming and quiescent galaxies, respectively. All kinematic components in poststarburst galaxies except LEGA-C~210541~A have among the highest \othreea/H$\beta$ for galaxies of similar stellar masses, suggesting that the AGN is the source of the ionization. }
	\label{fig:mex}
\end{figure}

To unveil the subtle emission lines present in poststarburst galaxies, we subtract best-fit stellar continuum models from the 1D spectra, as demonstrated in Figure~\ref{fig:em}. Despite their weak H$\beta$ emission, our poststarburst galaxies exhibit well-detected lines in a few instances due to the high signal-to-noise ratios. Along with H$\beta$, we detect the presence of [\ion{O}{2}]$\lambda$3727,3729 and [\ion{O}{3}]$\lambda$4959,5007 in the majority of our sample galaxies. Additionally, LEGA-C~209377 exhibits [\ion{Ne}{3}]3869 and LEGA-C~210541 shows a tentative detection. The shapes of these emission lines, particularly that of \othreea, show clear asymmetry and hint at the presence of multiple kinematic components, including broad, blue-shifted emission.

\begin{figure*}
	\includegraphics[width=0.95\textwidth]{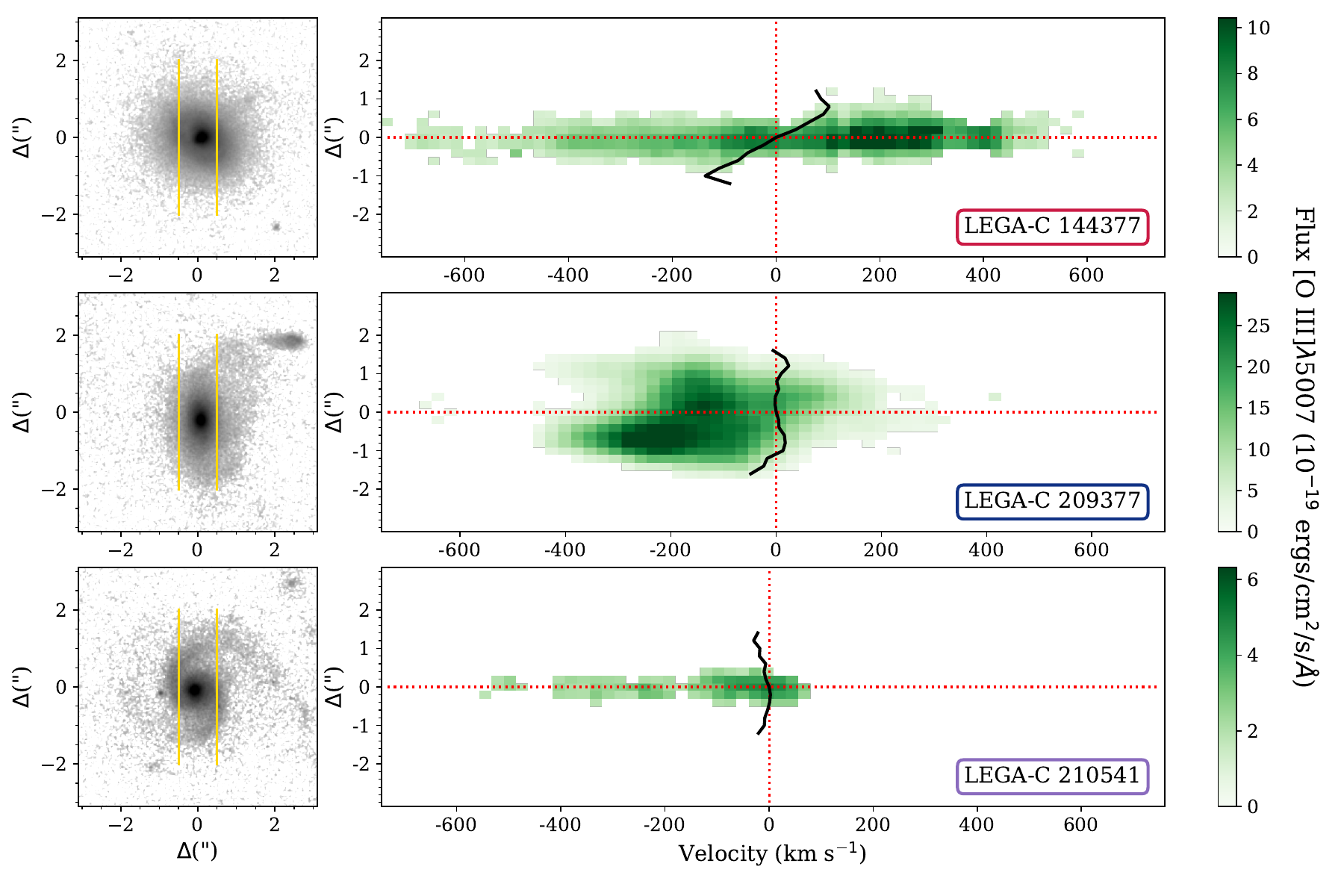}
	\caption{The position-velocity diagrams of the \othreea\ emission. The systematic velocity is set by the stellar velocity at the central pixel. The black solid lines are the stellar velocities. The kinematics of the \othreea\ gas is detached from the stars. The velocities of the wind can be up to a few hundred \kms. In LEGA-C~144377 and LEGA-C~210541, the gas is concentrated at galaxy centers.}
	\label{fig:pv}
\end{figure*}

In Figure~\ref{fig:pv}, we present the [\ion{O}{3}]$\lambda5007$ position-velocity (PV) diagrams along the slits of LEGA-C~144377, 209377, and 210541. By fitting stellar population synthesis models to each row of the 2D spectra, we subtract out the stellar continua similar to Figure~\ref{fig:em}. Only pixels with $S/N \ge 2$ are presented, and for comparison, the stellar velocities are also plotted.

The [\ion{O}{3}] emission of LEGA-C~144377 and LEGA-C~210541 appear to be spatially unresolved, with flux distributions that can be modeled as Gaussian distributions, centered on the galaxy centers, and with FWHM of approximately $\sim0\farcs8$. In contrast, LEGA-C~209377 exhibits multiple kinematic components on its PV diagram.

We employ a multi-Gaussian approach to model the 1D spectra, fitting \othreea\ and H$\beta$ simultaneously while assuming that the two lines are produced by the same kinematic components, sharing the same velocity and FWHM. To mitigate potential issues caused by imperfect continuum subtraction, we exclude the weaker \othreeb\ from the fitting process. The wavelength ranges for fitting are carefully selected based on visual inspection of the line emission. In the case of LEGA-C~144377 and LEGA-C~210541, we find that 2 kinematic components are sufficient, while LEGA-C~209377 requires 3 components, as determined by visual inspection of the fitting results and the PV diagram. The sums and individual Gaussians are displayed in Figure~\ref{fig:em}. LEGA-C~260163 exhibits clear [\ion{O}{2}] emission but no H$\beta$ emission. We attempt to model the [\ion{O}{2}] doublet with Gaussians, however, the uncertainties are too large to yield a convincing model. Similarly, we are unable to obtain a satisfactory model for the [\ion{O}{2}] of LEGA-C~209377 either, likely due to its complex velocity structure as shown by H$\beta$ and \othreea. The central velocities and FWHM as well as fluxes of each kinematic component are listed in Table~\ref{tab:em}. For the [\ion{O}{2}] emission that cannot fit well, we determined its flux by summing over $\pm800$ \kms. We also calculate the total fluxes from the direct sum of H$\beta$ and \othreea\ of all galaxies and the results are consistent with Gaussian models.

We find that the line-of-sight velocities of many kinematic components are $\sim200$~\kms\ different from those of the stellar bodies. Some of them have line widths of $\gtrsim 500$~\kms. The line emission is thus unlikely powered by star-formation activities. Meanwhiles, the \othreea/H$\beta$ flux ratios of most kinematic components rank among the highest in the LEGA-C sample (Figure~\ref{fig:mex}). For LEGA-C~209377, although the fluxes of individual components may be uncertain, the total fluxes determined through direct sum are reliable measurements (Tabel~\ref{tab:em}). Based on their \othreea/H$\beta$ flux ratios and stellar masses, all components except LEGA-C~210541~A would be classified as AGN-driven \citep{jun14}.

LEGA-C~210541~B and the integrate fluxes of LEGA-C~209377 have \othreea/H$\beta$ ratios $\gtrsim10$, which cannot be produced by shock-ionization \citep[$\lesssim2.5$, see, e.g.,][]{ric11} or be attributed to low-ionization nuclear emission-line regions (LINERs) \citep{kew06}. The emission in LEGA-C~144377 has line widths of $\sim500-1000$~\kms, which is also unlikely powered by old stars or shocks \citep{ho14}. Therefore, AGN is likely the main driver of line emission in these galaxies. However, no X-ray counterparts are found in public X-ray source catalogs \citep{cap09,civ16,mar16} and their \textit{Spitzer} IRAC mid-IR colors \citep{ste12} do not classify them as AGN either (S. Velvalcke, private communication).

\subsection{Kinematics of CO}
\label{sec:kin_CO}
\begin{figure*}
	\includegraphics[width=0.95\textwidth]{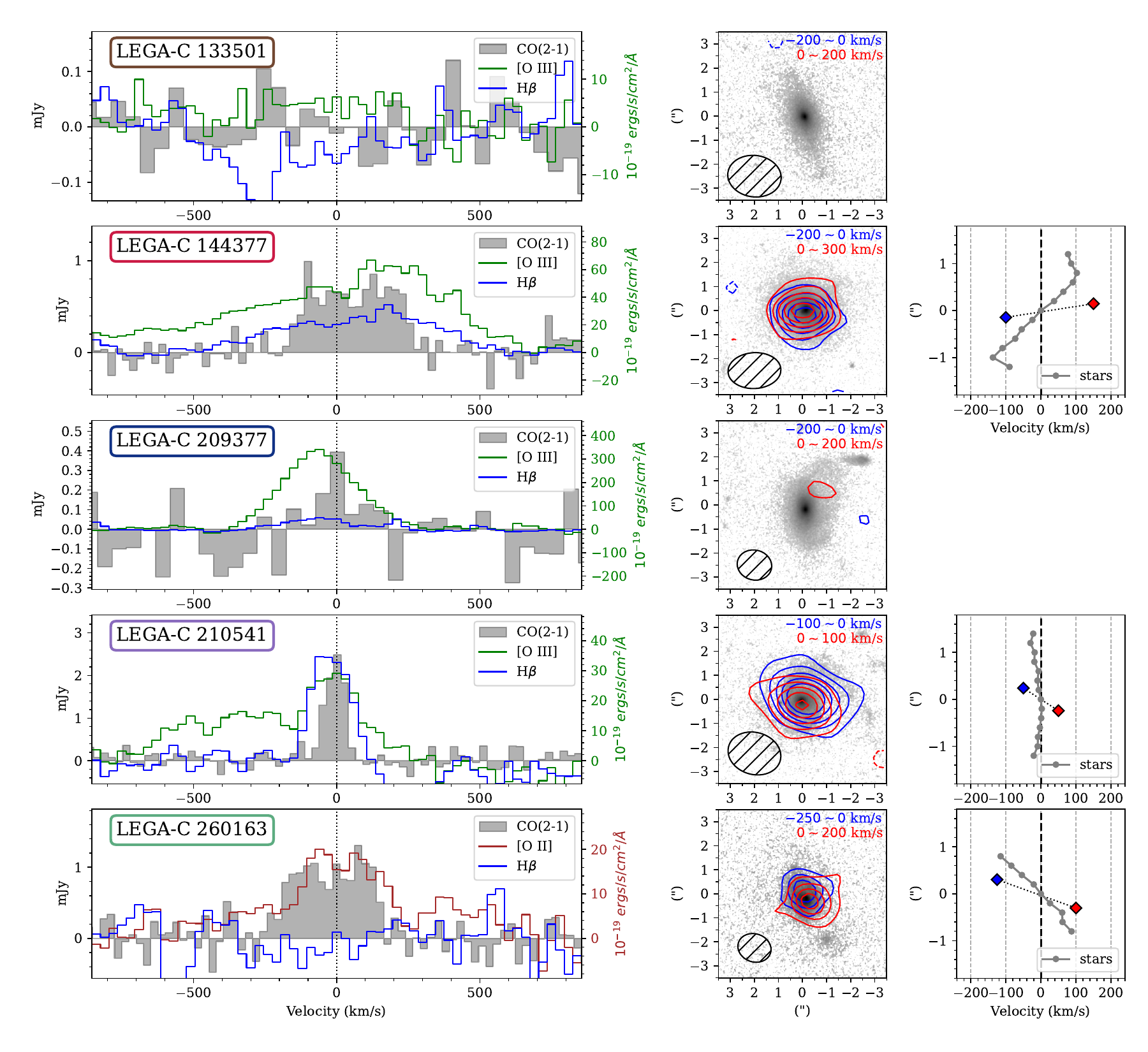}
	\caption{Kinematics of CO, ionized gas, and stars. \textit{First column:} The spectra of CO(2-1) (gray), H$\beta$ (blue), [\ion{O}{3}] (green), and [\ion{O}{2}] (brown) of poststarburst galaxies. The CO spectra are extracted in a $4”\times4”$ squared region located at the optical centers of galaxies. The spectra of LEGA-C 144377, 210541, and 260163 are binned to $\sim25$\kms. The rest two are binned to $\sim50$\kms\ for presentation. The kinematics of CO(2-1) and ionized gas are decoupled. \textit{Middle column:} The spatial distributions of red- and blue-shifted CO emission overplotted on the \textit{HST} F814W images. Contours are every 3$\sigma$. The centroids of blue- and redshifted CO emission show small displacements. \textit{Right column:} Comparison of the stellar (gray circles) and CO (colored squares) velocities. The positions of CO are the centroids measured from Gaussian fits (see main text) projected in the N-S direction. The velocity gradients of CO is in the same direction as the stars but stronger.}
	\label{fig:CO}
\end{figure*}

The first column of Figure~\ref{fig:CO} presents the CO spectra, integrated over an area of 2" from the centers of  and binned in velocity for clarity. These spectra exhibit double-horn profiles for LEGA-C~144377 and LEGA-C~260163, while LEGA-C~210541 shows a centrally-peaked profile. The tentative detection of LEGA-C~209377 is observed at $\sim0$\kms. Such double-horn and singly-peaked profiles have been previously noted in poststarburst galaxies at similar redshifts \citep{sue17,bez22}.

Furthermore, the overplotted ionized emission line profiles reveal that for LEGA-C~144377, 209377, and 260163, the velocity distribution of the CO does not align with those of H$\beta$, [\ion{O}{3}], or [\ion{O}{2}]. Similarly, LEGA-C~210541 presents a broad blueshift ionized gas component that lacks a corresponding CO counterpart. The CO has a similar central velocity to that of the narrow component LEGA-C~210541~A but with a smaller FWHM ($\sim100$~\kms\ compared to $\sim150$~\kms). LEGA-C~210541~A is also the only component with the \othreea/H$\beta$ flux ratio consistent with being powered by star-formation activities.

The second column of Figure~\ref{fig:CO} presents maps of red- and blue-shifted CO emission, with the velocity ranges for integration chosen by visual inspection. We detect slight offsets between the red and blue components and determine their centroids through fitting with either Gaussian or point source models in both image and visibility spaces. The separations between the red and blue components ranging from $0\farcs25$ to $0\farcs5$ and all methods yield consistent results.

In the third column of Figure~\ref{fig:CO}, we compare the velocity gradients of CO and stars along the N-S direction, acknowledging the vastly different angular resolutions and measurement types of the optical and sub-mm data. We observe that the velocity gradients of stars and CO in LEGA-C~144377 and LEGA-C~260163 are in the same direction, but the CO appears to have stronger gradients. Given the worse angular resolution of the ALMA data, thus the more serious beam smearing effect, the higher CO velocity gradients are likely real. For the flat galaxy LEGA-C~260163, the CO velocity gradient is clearly along the photometric major axis of the stellar light, potentially indicating compact disks co-rotating with the stars at the central $\lesssim0\farcs5$.

\section{Star-formation rate}
\label{sec:sfr}

By definition, poststarburst galaxies are galaxies with low SFRs. Star-formation tracers from poststarburst galaxies thus should be faint by nature. The 5 poststarburst galaxies are rare cases where multiple SFR indicators are available together. We see that different SFR indicators can disagree with each other by an order of magnitude or more (Table~\ref{tab:sfr}). Overall, $SFR_{UV+IR}$ and $SFR_{3GHz}$ are higher than $SFR_{SED}$ and $SFR_{H\beta}$.

All the SFRs have their own shortcomings. First of all, the fiducial calibration of the FIR-based SFR assumes a constant SFR in the past $\sim$100~Myrs. Based on our models, this assumption clearly cannot apply to poststarburst galaxies. \citet{hay14} post-processed hydrodynamical simulations with dust radiative transfer and calculated the SED of galaxies during the galaxy merging processes. They found that the FIR luminosity can overestimate the SFR by more than 10 times in the poststarburst phase because the dust is mainly heated by the prodigious amount of radiation from stars with ages of hundreds of Myr.

\begin{figure*}
\gridline{\fig{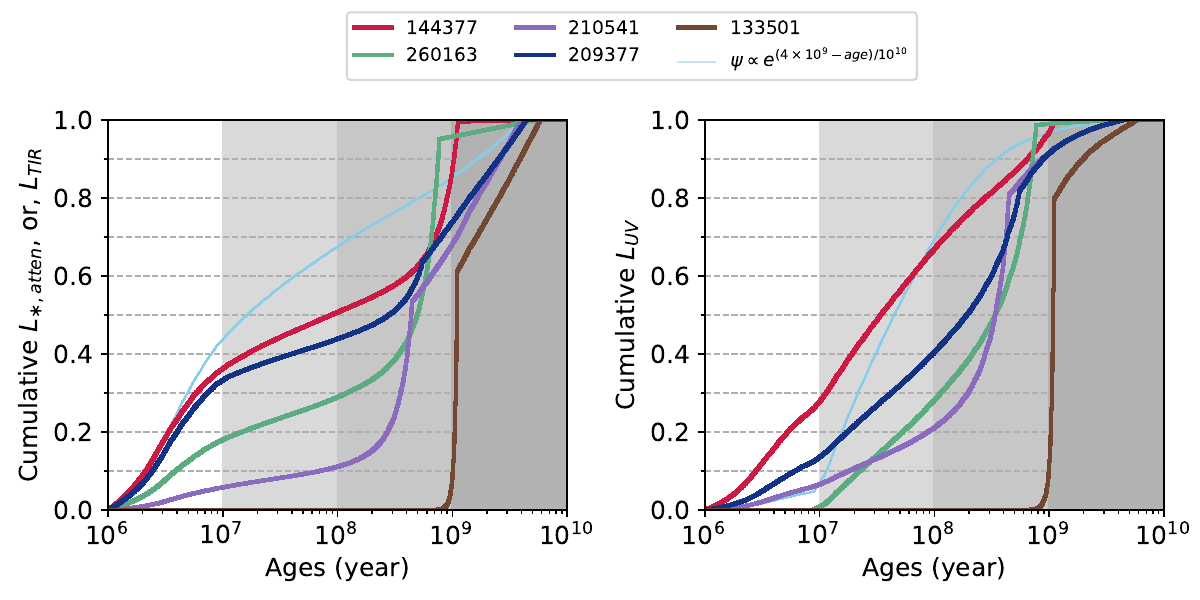}{0.65\textwidth}{(a)}
          \fig{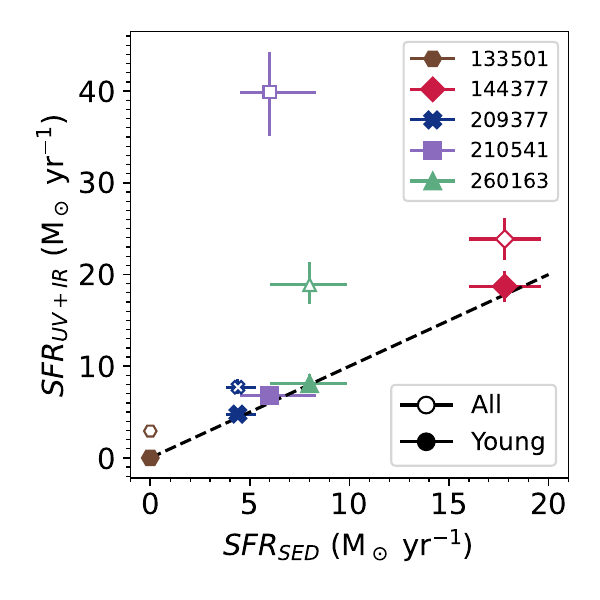}{0.32\textwidth}{(b)} }
\caption{(a) The cumulative distribution of attenuated stellar radiation and UV luminosities as a function of stellar ages of poststarburst galaxies. The curves are calculated based on the best-fit SFHs and dust attenuation. The attenuated light then re-emitted in the IR. About 70\% of the $L_{TIR}$ of a star-forming galaxy (light blue) is heated by stars younger than 100~Myrs. The contribution from stars of $<100$~Myr can be much lower in poststarburst galaxies. Similarly, a larger fraction of the UV radiation of poststarburst galaxies is contributed by older stars compared to star-forming galaxies. The radiation from old stars results in overestimates of SFRs based on FIR and UV luminosities. (b) The contamination from old stars to the $SFR_{UV+IR}$. The filled symbols show the $SFR_{UV+IR}$ after excluding the contribution from stars older than 100~Myrs (see main text). The new $SFR$ are much lower than original $SFR_{UV+IR}$ and more consistent with $SFR_{SED}$. \label{fig:cumlum}}
\end{figure*}

Based on the model SFH and dust attenuation (Section~\ref{sec:pipes}), we calculate how much the radiation is attenuated by dust for stars of different ages. This quantity can be interpreted as the contribution to dust heating from different ages of stars under the common energy balance assumption, which is adopted in our fitting scheme. Figure~\ref{fig:cumlum}a shows the normalized cumulative distributions of attenuated stellar radiation as a function of stellar ages. The curves also inform us of the main stellar population that heats up the dust.

We show a model star-forming galaxy with $SFR(t) \propto \exp(-t/10\ Gyr) $, $Z_\odot = 1$, $A_v=1.5$, $n=0.8$, and $\eta=2$ for comparison. For the model star-forming galaxy, stars younger than 100~Myrs contribute to $\sim67\%$ of the dust heating. On the other hand, for poststarburst galaxies which have significantly larger $SFR_{UV+IR}$ than $SFR_{SED}$ (LEGA-C~133501, 210541, and 260163), the majority ($\sim60\%$) of the dust heating is provided by stars with ages between 100~Myrs and 1~Gyrs. Stars younger than 100~Myrs only contribute $\lesssim 30\%$, even $0\%$, of the energy. Adopting the fiducial FIR-based SFR calibration can potentially overestimate the SFR by more than an order of magnitude if the SFR was rapidly declining in the recent past. Similarily, the large number of old stars also contributes significantly to the UV radiation (Figure~\ref{fig:cumlum}a). 

With the SFHs, we can attempt to quantify how much the $SFR_{UV+IR}$ is overestimated for poststarburst galaxies. We have computed the $SFR_{UV+IR}$ as $\Psi[M_\odot\ \mbox{yr}^{-1}] = 1.09\times10^{-10} (L_{IR} + 2.2 L_{UV}) [L_\odot]$. Taking the model star-forming galaxy in Figure~\ref{fig:cumlum} as the baseline, $67\%$ and $70\%$ of the total IR and UV radiation is contributed by stars younger than 100~Myrs, respectively. Therefore, if we use only the contribution from young stars to estimate the SFR, we can rewrite the equation for the model star-forming galaxy as $\Psi[M_\odot\ \mbox{yr}^{-1}] = (1.63\times10^{-10} L_{IR_{<100 Myr}} +  3.46\times10^{-10} L_{UV_{<100 Myr}}) [L_\odot]$, where $L_{IR_{<100 Myr}}$ and $L_{UV_{<100 Myr}}$ are the luminosities from stars younger than 100~Myrs. We then apply this new `calibration' to poststarburst galaxies, plug in $L_{UV_{<100 Myr}}$ and $L_{IR_{<100 Myr}}$ based on the best-fit models. We find the new $SFR_{UV+IR_{<100 Myr}}$ is consistent with $SFR_{SED}$ (Figure~\ref{fig:cumlum}b). This result illustrates that the commonly-used calibration can overestimate the SFRs of poststarburst galaxies due to their peculiar SFHs. 

In addition, as AGN is present from the optical emission lines  (Figure~\ref{fig:mex}), we should not ignore the potential contributions to SFR tracers. Not only H$\beta$ emission is powered by AGN, but also IR and radio continua. Therefore, $SFR_{H\beta}$, $SFR_{UV+IR}$, and $SFR_{3GHz}$ may all overestimate the true SFR due to the contribution from AGN. The compact 3~GHz emission is morphologically consistent with being powered by AGN. On the other hand, the SPS models fit the optical spectra well (Figure~\ref{fig:fit}). There is no need to include a featureless continuum component. Therefore, we consider the SFRs from fitting the spectra are not contaminated by AGN and adopt $SFR_{SED}$ as our fiducial $SFR$.

Nevertheless, $SFR_{SED}$ should also be used with caution. Firstly, the posterior SFHs can be biased by the choice of priors \citep{car19a,lej19a,sue22}. Moreover, if poststarburst galaxies are indeed the descendant of dusty, compact central starbursts, then the youngest stars in poststarburst galaxies may still be deeply embedded in dust clouds \citep{pog00,koc11}. The attenuation at the optical wavelength in some poststarburst regions in a few local galaxies is argued to be up to thousands of magnitudes \citep{sme22}, therefore, optical and NIR light would not probe deep in the dust clouds and underestimates the true attenuation and SFRs. \citet{sme18} proposed to use MIR nebular neon emission lines, [\ion{Ne}{2}] at 12.8$\mu$m and [\ion{Ne}{3}] at 15.6$\mu$m, to avoid the issue of high attenuation \citep{ho07}. With the MIR nebular emission lines, the problem on SFRs may be solvable with JWST up to $z\sim0.8$. Given that SFRs are dominated by systematic uncertainties, there is still much work to be done to verify their low SFEs.

\section{Molecular gas content and star formation}
\label{sec:gas}

\begin{figure*}
    \includegraphics[width=0.95\textwidth]{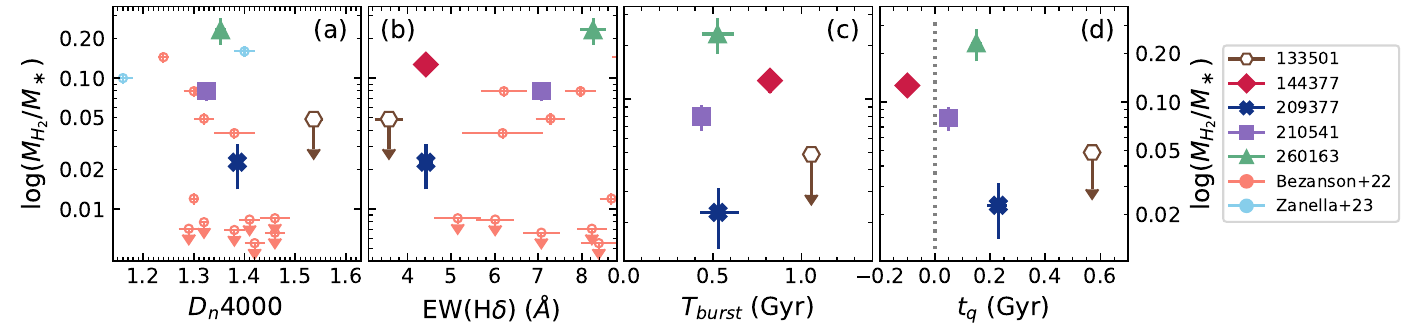}
    \caption{The correlations between molecular gas fraction and \dn, \ewhd, look-back time of the onset of the burst, and the look-back time when galaxies became quiescent. Poststarburst galaxies at similar redshifts with CO measurements in \citet{bez22} ($z\sim0.6$) and \citet{zan23} ($z\sim1.2$) are also plotted.\label{fig:fgas}}
\end{figure*}

\begin{figure*}
	\includegraphics[width=0.95\textwidth]{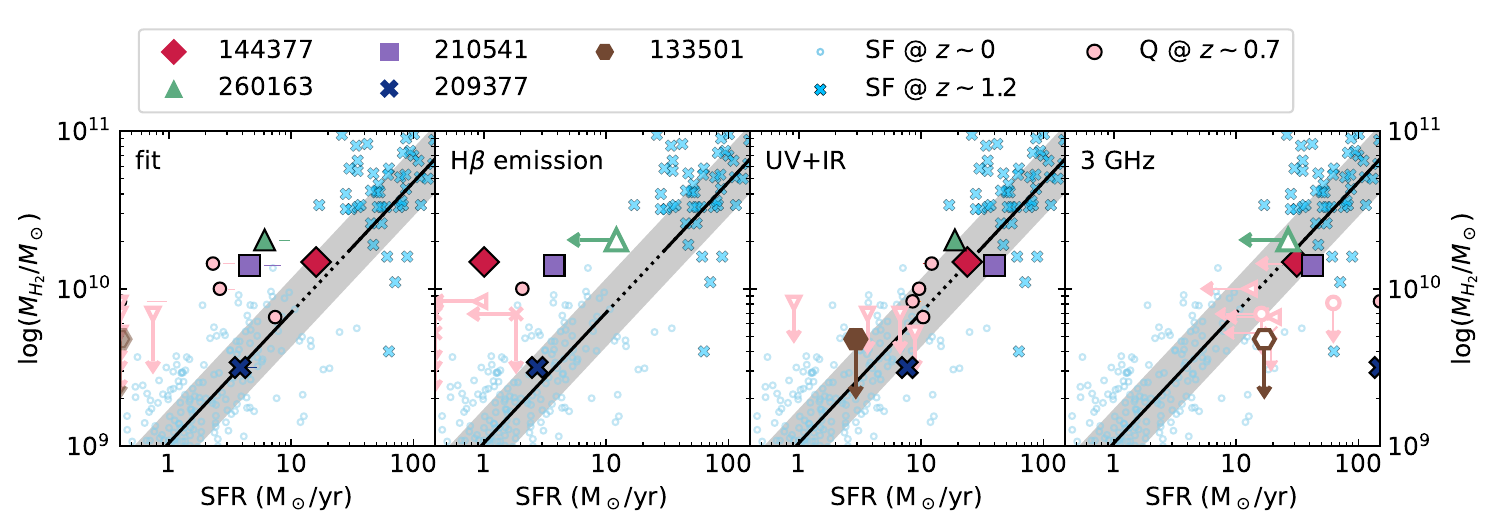}
	\caption{The $M_{H2}/M_\odot$ and SFRs of poststarburst galaxies. Each panel shows the SFR measured by different methods. Poststarburst galaxies with SFRs beyond the displayed ranges are placed at the edges of the panels. Pink circles are quiescent galaxies in \citet{spi18}, with SFRs measured by the same methods. Blue crosses and circles are star-forming galaxies at $z\sim1.2$ from the PHIBSS \citep{tac13} and $z\sim0$ from the xCOLDGASS \citep{sai11}, for which the SFRs and $M_{H2}$ are taken from the papers and adopting an $\alpha_{CO}=4$. The solid line is the scaling relation between SFR and $M_{H2}$ fitted with the PHIBSSS and xCOLDGASS data and the shaded area shows a 0.2~dex scatter. SFRs of poststarburst galaxies from different methods can differ by an order of magnitude. Adopting the more reliable $SFR_{SED}$, poststarburst galaxies have lower SFEs than typical star-forming galaxies. \label{fig:co_sfr}}
\end{figure*}

Figure~\ref{fig:fgas} shows the correlation between $f_{H_2}$ and age-related parameters of our poststarburst galaxies. Samples at similar redshifts \citep{bez22,zan23} are also plotted. Although caution must be taken in drawing any definitive conclusions from such a small sample size. Nevertheless, there appears to be a tentative trend indicating that poststarburst that became quiescent earlier exhibit lower $f_{H_2}$. The three galaxies with a clear detection of CO(2-1) also have lower $t_q$, indicating a more recent quiescent state (LEGA-C~144377, 201541, and 260163). This trend is consistent with the findings of \citet{bez22}, who recently measured the molecular gas and SFH of massive poststarburst galaxies at similar redshifts, and found that while poststarburst galaxies can initially contain molecular gas, it depletes rapidly within $\sim100$~Myr. The age-sensitive \dn\ reveal a consistent picture in such that there are no poststarburst galaxy with CO detection and with $D_n4000 \gtrsim 1.4$ (Figure~\ref{fig:fgas}).

We compare the molecular gas masses and SFRs of poststarburst galaxies (from Table~\ref{tab:sfr}) to those of normal star-forming galaxies (Figure~\ref{fig:co_sfr}). As a point of reference, we include star-forming galaxies at redshifts $z=1-1.3$ from the PHIBSS survey \citep[blue crosses,][]{tac13} and local massive star-forming galaxies from the xCOLDGASS sample \citep[blue circles,][]{sai11}. We also show the measurements of 8 massive quiescent galaxies from \citet{spi18} for comparison. The molecular gas masses have been adjusted to $\alpha_{CO}=4.0$ for consistency, and the SFRs are extracted from the papers.

We adopt $SFR_{SED}$ as our primary measurement of star formation rate, but we present the other $SFRs$ for completeness. We find that poststarburst galaxies exhibit similar or even higher molecular gas fractions than normal star-forming galaxies. Furthermore, when we adopt $SFR_{H\beta}$, we observe a larger deviation from the scaling relation. Due to the contamination of AGN, we anticipate that the true deviation is even larger.

These deviations suggest that poststarburst galaxies may have several times lower SFEs than normal star-forming galaxies. While they still possess a considerable amount of molecular gas, there should be some physical processes that suppress star formation, preventing the gas from converting to stars.

Interestingly, poststarburst galaxies fall roughly on the scaling relation in the 3rd and 4th panels. Had we relied on $SFR_{UV+IR}$ or $SFR_{3GHz}$ as proxies for SFR, our conclusion would have been vastly different. Nonetheless, we have established that $SFR_{UV+IR}$ tends to overestimate SFRs (Section~\ref{sec:sfr}). The same trend of overestimation has been observed in quiescent galaxies, where $SFR_{UV+IR}$ exceeds $SFR_{SED}$ and $SFR_{H\beta}$ systematically. This trend can be attributed to the fact that most of the light that heats up dust in these galaxies emanates from old stars.
For $SFR_{3GHz}$, if the 3~GHz continuum traces the star formation activities, we should expect that the 3~GHz continuum and the CO(2-1) emission have similar sizes. LEGA-C~210541 gives a strong piece of counter-evidence in such that its CO(2-1) emission extends over $\sim1\farcs3$ ($\sim10$~kpc) while the 3~GHz emission is unresolved under a $0\farcs75$ beam. Assuming a typical SFE and $SFR = SFR_{SED}$, the expected 3~GHz flux is below the current detection limit, in line with the fact that no 3~GHz source of $\sim1"$ is detected. The 3rd and 4th panels hence highlight the non-trivial task of accurately determining SFRs in poststarburst galaxies - a crucial aspect of understanding quenching.

\section{Discussion}
\label{sec:dis}

The diverse high-quality data provide us a rare opportunity to look simultaneously into the stars, gas, and star-formation of distant poststarburst galaxies. In this section, we discuss the implication of quenching from the multi-wavelength data.

We note that galaxies in this study do not reside in massive clusters. Galaxy clusters also play a role in triggering starburst and quenching at similar redshifts, through mechanisms that only operate in cluster environments \citep{muz12,wu14,mat21} but investigating these cluster-specific mechanisms is beyond the scope of this paper.

\subsection{Implication to the quenching process}

The poststarburst galaxies may be a remnant of the collision or merging of gas-rich galaxies that induced gas inflows, triggered intense star formation and AGN, and also fueled outflows that brought negative feedback \citep{bar96,hop06}. According to simulations, the poststarburst phase may last from a few hundred million years to nearly a billion years after the merging event \citep{bek05,sny11}. The disturbed stellar morphologies may be visible for a similar time frame of about half a billion years \citep{lot08,lot10a,lot10b}.

Our modeling suggests that these poststarburst galaxies experienced a starburst event several hundred million years ago and then underwent rapid quenching. Out of the five galaxies we observed, four exhibit disturbed morphologies as seen through either \textit{HST} or HSC imaging. The high incident rate of disturbed features is consistent with the observation from $\sim100$ massive poststarburst galaxies at similar redshift by \citet{ver22}. The residual star formation, if present, appears to be centralized around the galaxy cores (Figure~\ref{fig:cutout}). Molecular gas is also centrally concentrated in most cases. The multi-wavelength data point towards strong galaxy interaction as the primary cause.

From the SFHs and current gas mass, we can estimate the stellar and gas masses of the progenitors. The stellar mass of the progenitor is the current stellar mass minus the mass formed in the burst, taking into account the mass returned into the interstellar medium (ISM) towards the end of the stellar evolution. The gas mass of the progenitor is the sum of the current gas mass and the stellar mass formed in the burst. Using the values derived in Section~\ref{sec:pipes} and Section~\ref{sec:alma}, we find a wide range of initial stellar masses and gas fractions of progenitors. Despite the present-day stellar masses are all $1-2 \times 10^{11} M_\odot$, the stellar masses before bursts range from $\sim 2 \times 10^{9} M_\odot$ to $\sim10^{11} M_\odot$ in our sample. Moreover, the molecular gas fraction $M_{H_2}/M_\ast$ ranges from $\sim0.3$ for the high-mass progenitors to $\sim50$ for the lowest-mass progenitor. In this calculation, we have neglected the possible contribution of atomic hydrogen. For local poststarburst galaxies with $M_\ast > 10^{10} M_\odot$, \citet{zwa13} shows that $M_{HI}/M_\odot$ is generally a few percent. Since there is no information for larger distances, we apply the local value. Also, we have not accounted for potential mass loss caused by tidal disruption or outflows during the process, as observational constraints are lacking. However, the off-center detection of CO in LEGA-C 209377 and a case study by \citet{spi22} show that gas stripping occurs. Therefore, the gas fraction we derive may be considered as a lower limit.

Our observations suggest that galaxies with $M_\ast \sim 10^{11} M_\odot$ and $M_{H_2}/M_\ast \sim 0.3$ could be progenitors of poststarburst galaxies. Such a molecular gas fraction is fairly typical for star-forming galaxies at $z\sim1$ \citep{tac13,wan22}. This result indicates that rapid elevation and then truncation of SFR could happen in normal star-forming galaxies and may not require a particularly gas-rich progenitor. For the other extreme, there are currently fewer measurements of molecular gas contents for lower-mass galaxies with $M_\ast \sim 10^9 M_\odot$ but galaxies with $M_{H_2}/M_\ast>10$ have been found however very rare at $z\sim1$ \citep{liu19}. These extremely gas-rich systems may become the most extreme poststarburst galaxies at later times.

It will be intriguing to see whether hydrodynamical simulations are able to reproduce these observations; from the inferred wide ranges of initial stellar and gas masses, produce starbursts with such rapid and extreme bursts with 100 Myr e-folding times, forming several $10^{10} M_\odot$ stars, and match the morphologies at the similar time after the burst as constrained by the data. Answering these questions with accuracy could shed light on the mechanisms driving poststarburst galaxies and the role of galaxy interactions and feedback in quenching their star formation.

\subsection{The role of AGN}
\label{sec:dis_agn}

In the quenching process, the role of AGN is considered to be crucial due to its energy and momentum output. Despite being frequently associated with the quenching of star formation in poststarburst galaxies, many poststarburst galaxies do not exhibit bright radio and X-ray emission or distinct mid-IR colors \citep{nie12,meu17}. Only one poststarburst galaxy in this study shows luminous radio continuum emission, none is detected in X-ray, and none is classified as AGN based on the mid-IR color selection of \citet{ste12}. The low radio AGN fraction is consistent with the previous study of \citet{gre20}, who demonstrated that only 5\% of $\sim1000$ massive poststarburst galaxies at $z\sim0.7$ exhibit $L_{3GHz} \gtrsim 10^{24}$~W~Hz$^{-1}$. Notably, the radio and X-ray surveys conducted in the COSMOS field are among the deepest blind surveys to date. Thus, the lack of radio and X-ray detections unequivocally indicates the low emission levels of poststarburst galaxies at these wavelengths.

In contrast, when \othreea\ falls within the spectral range, three out of 4 poststarburst galaxies exhibit \othreea\ gas flows powered by AGN (Tabel~\ref{tab:em} and Figure~\ref{fig:mex}). This high incidence rate of AGN-driven \othreea\ emission stands in contrast to the $\sim5\%$ incidence rate reported by \citet{gre20}. Our high incidence rate is likely attributable to the ultra-deep LEGA-C spectra, which allowed us to precisely model the stellar continua and discern weak emission lines. Our results suggest that AGN may be more prevalent than previously assumed.

Nevertheless, the impact of the AGN-driven wind on CO gas is unclear. Evidence of outflowing CO gas has been reported \citep{spi22}, but in our sample, there are no indications of disturbed morphology or high-velocity components at our current angular resolution and sensitivity, despite the detection of AGN-driven ionized gas. While it is possible that AGN activity plays a role in quenching star formation, our observations suggest that any significant impact on the molecular gas likely occurs prior to the phase identified in our study. Furthermore, it is possible that the AGN has not fully cleared or destroyed the gas reservoir thus residual star formation is allowed. This conclusion is consistent with local integral field unit (IFU) observations of poststarburst galaxies, showing residual star formation at sub-kpc scales in their centers \citep{wu21a}. In addition, a higher incidence rate of AGN in poststarburst galaxies with younger stellar ages, as found by \citet{gre20}, suggests that the peak AGN activity may occur during an earlier phase of the quenching process.

\section{Summary and Future Work}
\label{sec:sum}

We present a comprehensive study of 5 massive poststarburst galaxies with $M_\odot \gtrsim 10^{11} M_\odot$ at $z\sim0.7$, utilizing multi-wavelength observations to trace the stars, gas, and current and past star formation activities.

The ultra-deep spectra from the LEGA-C survey give us good constraints on the SFHs. We find that these galaxies underwent bursts of star formation several hundred Myr ago, with a decay time scale of approximately 100~Myr. Prior to the burst, these massive poststarburst galaxies were heterogeneous, with stellar masses ranging from $2\times10^9 M_\odot$ to $>10^{11} M_\odot$, and with gas masses up to 50 times the stellar mass for the lowest-mass progenitor to $\sim30\%$ of the stellar mass for the high-mass progenitors. Our deep spectra also reveal that weak AGN-driven [\ion{O}{3}] outflows are common in poststarburst galaxies.

Examining high-resolution images from HST and deep images from HSC, we find in our sample of 5 galaxies, 4 of them have low surface brightness tidal features that extend over tens of kpc. In addition, 4 of them have a companion galaxy within $\sim100$~kpc, with $K_s$-band flux greater than 20\% of the poststarburst galaxy, and at a similar redshift.

Some poststarburst galaxies possess CO gas of $\sim10^{10} M_\odot$ at the galaxy centers, with intrinsic sizes of $>1\arcsec$, or $\sim10$~kpc. In contrast, neither the 3~GHz continuum image from VLA nor the H$\beta$ emission from LEGA-C slit spectra suggest extended star-forming regions. Both of these observations have angular resolutions of $\sim0\farcs8$. Our results suggest that the star-forming regions are more compact than the CO gas, and the star-formation activities are suppressed. Comparing the SFRs derived from SED fitting and the molecular contents leads us to the same conclusion; poststarburst galaxies have lower SFEs than star-forming galaxies.

Meanwhiles, we find that the SFRs obtained through UV and IR continuum and 3~GHz continuum tend to overestimate the SFR of poststarburst galaxies. This is due to the fact that, unlike normal star-forming galaxies, stars older than 100~Myr are the primary source of dust heating and UV radiation in poststarburst galaxies, rendering IR and UV continua inadequate indicators for star formation. Moreover, both IR and 3~GHz continuum flux may reflect not only star formation but also AGN activity. Despite that the AGN-driven ionized outflows are prevalent, we do not find CO outflow at current angular resolutions and sensitivities. While AGN may contribute to quenching as indicated by many studies, they do not entirely clear out or destroy the cold gas reservoir.

The multi-wavelength observations of the 5 galaxies in our study provide compelling evidence that quenching is linked to interactions between galaxies, which can trigger centrally-concentrated starbursts followed by a rapid truncation of star formation. AGN is common in this phase. This process is not restricted to gas-rich, low-stellar-mass galaxies, but can also occur in high-stellar-mass galaxies where the gas mass is only a small fraction of the stellar mass like typical massive star-forming galaxies at $z\sim1$. When combined with the SFHs derived from deep optical spectra, our findings, offer valuable insights into the changes that take place in galaxy properties during the quenching process and can be used to refine simulations of galaxy formation and evolution. This work is a step forward to understanding the mechanisms that shut down star formation in distant galaxies.

\begin{acknowledgments}
PFW thanks Dr. Chen Chien-Chou and Dr. Sharon Meidt for fruitful discussions. PFW acknowledges funding through the National Science and Technology Council grant 111-2112-M-002-048-MY2 and National Taiwan University grant NTU 112L7318 and NTU 112L7439.
FDE acknowledges funding through the ERC Advanced grant 695671 ``QUENCH’’ and support by the Science and Technology Facilities Council (STFC). AG acknowledges support from INAF-Minigrant-2022 "LEGA-C" 1.05.12.04.01.
This paper makes use of the following ALMA data: ADS/JAO.ALMA\#2019.1.00702.S. ALMA is a partnership of ESO (representing its member states), NSF (USA) and NINS (Japan), together with NRC (Canada), MOST and ASIAA (Taiwan), and KASI (Republic of Korea), in cooperation with the Republic of Chile. The Joint ALMA Observatory is operated by ESO, AUI/NRAO and NAOJ.

\end{acknowledgments}

\facilities{ALMA, HST (ACS), Sptizer, Subaru, VLA, VLT:Melipal (VIMOS)}

\software{astropy \citep{astropy13,astropy18}, BAGPIPES \citep{car18}, pPXF \citep{cap04,cap17}, Matplotlib \citep{hun07}}

\bibliography{LEGAC}{}

\begin{thebibliography}{}
\expandafter\ifx\csname natexlab\endcsname\relax\def\natexlab#1{#1}\fi
\providecommand{\url}[1]{\href{#1}{#1}}
\providecommand{\dodoi}[1]{doi:~\href{http://doi.org/#1}{\nolinkurl{#1}}}
\providecommand{\doeprint}[1]{\href{http://ascl.net/#1}{\nolinkurl{http://ascl.net/#1}}}
\providecommand{\doarXiv}[1]{\href{https://arxiv.org/abs/#1}{\nolinkurl{https://arxiv.org/abs/#1}}}

\bibitem[{{Asayama} {et~al.}(2014){Asayama}, {Takahashi}, {Kubo}, {Ito},
  {Inata}, {Suzuki}, {Wada}, {Soga}, {Kamada}, {Karatsu}, {Fujii}, {Obuchi},
  {Kawashima}, {Iwashita}, \& {Uzawa}}]{asa14}
{Asayama}, S., {Takahashi}, T., {Kubo}, K., {et~al.} 2014, \pasj, 66, 57,
  \dodoi{10.1093/pasj/psu026}

\bibitem[{{Astropy Collaboration} {et~al.}(2013){Astropy Collaboration},
  {Robitaille}, {Tollerud}, {Greenfield}, {Droettboom}, {Bray}, {Aldcroft},
  {Davis}, {Ginsburg}, {Price-Whelan}, {Kerzendorf}, {Conley}, {Crighton},
  {Barbary}, {Muna}, {Ferguson}, {Grollier}, {Parikh}, {Nair}, {Unther},
  {Deil}, {Woillez}, {Conseil}, {Kramer}, {Turner}, {Singer}, {Fox}, {Weaver},
  {Zabalza}, {Edwards}, {Azalee Bostroem}, {Burke}, {Casey}, {Crawford},
  {Dencheva}, {Ely}, {Jenness}, {Labrie}, {Lim}, {Pierfederici}, {Pontzen},
  {Ptak}, {Refsdal}, {Servillat}, \& {Streicher}}]{astropy13}
{Astropy Collaboration}, {Robitaille}, T.~P., {Tollerud}, E.~J., {et~al.} 2013,
  \aap, 558, A33, \dodoi{10.1051/0004-6361/201322068}

\bibitem[{{Balogh} {et~al.}(1999){Balogh}, {Morris}, {Yee}, {Carlberg}, \&
  {Ellingson}}]{bal99}
{Balogh}, M.~L., {Morris}, S.~L., {Yee}, H.~K.~C., {Carlberg}, R.~G., \&
  {Ellingson}, E. 1999, \apj, 527, 54, \dodoi{10.1086/308056}

\bibitem[{{Bari{\v s}i{\'c}} {et~al.}(2017){Bari{\v s}i{\'c}}, {van der Wel},
  {Bezanson}, {Pacifici}, {Noeske}, {Mu{\~n}oz-Mateos}, {Franx}, {Smol{\v
  c}i{\'c}}, {Bell}, {Brammer}, {Calhau}, {Chauk{\'e}}, {van Dokkum}, {van
  Houdt}, {Gallazzi}, {Labb{\'e}}, {Maseda}, {Muzzin}, {Sobral}, {Straatman},
  \& {Wu}}]{bar17}
{Bari{\v s}i{\'c}}, I., {van der Wel}, A., {Bezanson}, R., {et~al.} 2017, \apj,
  847, 72, \dodoi{10.3847/1538-4357/aa8768}

\bibitem[{{Barnes} \& {Hernquist}(1996)}]{bar96}
{Barnes}, J.~E., \& {Hernquist}, L. 1996, \apj, 471, 115,
  \dodoi{10.1086/177957}

\bibitem[{{Baron} {et~al.}(2022){Baron}, {Netzer}, {French}, {Lutz}, {Davies},
  \& {Prochaska}}]{bar22}
{Baron}, D., {Netzer}, H., {French}, K.~D., {et~al.} 2022, arXiv e-prints,
  arXiv:2204.11881, \dodoi{10.48550/arXiv.2204.11881}

\bibitem[{{Barro} {et~al.}(2013){Barro}, {Faber}, {P{\'e}rez-Gonz{\'a}lez},
  {Koo}, {Williams}, {Kocevski}, {Trump}, {Mozena}, {McGrath}, {van der Wel},
  {Wuyts}, {Bell}, {Croton}, {Ceverino}, {Dekel}, {Ashby}, {Cheung},
  {Ferguson}, {Fontana}, {Fang}, {Giavalisco}, {Grogin}, {Guo}, {Hathi},
  {Hopkins}, {Huang}, {Koekemoer}, {Kartaltepe}, {Lee}, {Newman}, {Porter},
  {Primack}, {Ryan}, {Rosario}, {Somerville}, {Salvato}, \& {Hsu}}]{barr13}
{Barro}, G., {Faber}, S.~M., {P{\'e}rez-Gonz{\'a}lez}, P.~G., {et~al.} 2013,
  \apj, 765, 104, \dodoi{10.1088/0004-637X/765/2/104}

\bibitem[{{Bekki} {et~al.}(2005){Bekki}, {Couch}, {Shioya}, \&
  {Vazdekis}}]{bek05}
{Bekki}, K., {Couch}, W.~J., {Shioya}, Y., \& {Vazdekis}, A. 2005, \mnras, 359,
  949, \dodoi{10.1111/j.1365-2966.2005.08932.x}

\bibitem[{{Bell}(2003)}]{bel03}
{Bell}, E.~F. 2003, \apj, 586, 794, \dodoi{10.1086/367829}

\bibitem[{{Bell} {et~al.}(2004){Bell}, {Wolf}, {Meisenheimer}, {Rix}, {Borch},
  {Dye}, {Kleinheinrich}, {Wisotzki}, \& {McIntosh}}]{bel04}
{Bell}, E.~F., {Wolf}, C., {Meisenheimer}, K., {et~al.} 2004, \apj, 608, 752,
  \dodoi{10.1086/420778}

\bibitem[{{Bell} {et~al.}(2005){Bell}, {Papovich}, {Wolf}, {Le Floc'h},
  {Caldwell}, {Barden}, {Egami}, {McIntosh}, {Meisenheimer},
  {P{\'e}rez-Gonz{\'a}lez}, {Rieke}, {Rieke}, {Rigby}, \& {Rix}}]{bel05}
{Bell}, E.~F., {Papovich}, C., {Wolf}, C., {et~al.} 2005, \apj, 625, 23,
  \dodoi{10.1086/429552}

\bibitem[{{Belli} {et~al.}(2019){Belli}, {Newman}, \& {Ellis}}]{bel19}
{Belli}, S., {Newman}, A.~B., \& {Ellis}, R.~S. 2019, \apj, 874, 17,
  \dodoi{10.3847/1538-4357/ab07af}

\bibitem[{{Best} {et~al.}(2005){Best}, {Kauffmann}, {Heckman}, {Brinchmann},
  {Charlot}, {Ivezi{\'c}}, \& {White}}]{bes05}
{Best}, P.~N., {Kauffmann}, G., {Heckman}, T.~M., {et~al.} 2005, \mnras, 362,
  25, \dodoi{10.1111/j.1365-2966.2005.09192.x}

\bibitem[{{Bezanson} {et~al.}(2022){Bezanson}, {Spilker}, {Suess}, {Setton},
  {Feldmann}, {Greene}, {Kriek}, {Narayanan}, \& {Verrico}}]{bez22}
{Bezanson}, R., {Spilker}, J.~S., {Suess}, K.~A., {et~al.} 2022, \apj, 925,
  153, \dodoi{10.3847/1538-4357/ac3dfa}

\bibitem[{{Bolatto} {et~al.}(2013){Bolatto}, {Wolfire}, \& {Leroy}}]{bol13}
{Bolatto}, A.~D., {Wolfire}, M., \& {Leroy}, A.~K. 2013, \araa, 51, 207,
  \dodoi{10.1146/annurev-astro-082812-140944}

\bibitem[{{Brinchmann} {et~al.}(2004){Brinchmann}, {Charlot}, {White},
  {Tremonti}, {Kauffmann}, {Heckman}, \& {Brinkmann}}]{bri04}
{Brinchmann}, J., {Charlot}, S., {White}, S.~D.~M., {et~al.} 2004, \mnras, 351,
  1151, \dodoi{10.1111/j.1365-2966.2004.07881.x}

\bibitem[{{Bruzual} \& {Charlot}(2003)}]{bc03}
{Bruzual}, G., \& {Charlot}, S. 2003, \mnras, 344, 1000,
  \dodoi{10.1046/j.1365-8711.2003.06897.x}

\bibitem[{{Cappellari}(2017)}]{cap17}
{Cappellari}, M. 2017, \mnras, 466, 798, \dodoi{10.1093/mnras/stw3020}

\bibitem[{{Cappellari} \& {Emsellem}(2004)}]{cap04}
{Cappellari}, M., \& {Emsellem}, E. 2004, \pasp, 116, 138,
  \dodoi{10.1086/381875}

\bibitem[{{Cappelluti} {et~al.}(2009){Cappelluti}, {Brusa}, {Hasinger},
  {Comastri}, {Zamorani}, {Finoguenov}, {Gilli}, {Puccetti}, {Miyaji},
  {Salvato}, {Vignali}, {Aldcroft}, {B{\"o}hringer}, {Brunner}, {Civano},
  {Elvis}, {Fiore}, {Fruscione}, {Griffiths}, {Guzzo}, {Iovino}, {Koekemoer},
  {Mainieri}, {Scoville}, {Shopbell}, {Silverman}, \& {Urry}}]{cap09}
{Cappelluti}, N., {Brusa}, M., {Hasinger}, G., {et~al.} 2009, \aap, 497, 635,
  \dodoi{10.1051/0004-6361/200810794}

\bibitem[{{Carnall}(2017)}]{car17}
{Carnall}, A.~C. 2017, arXiv e-prints, arXiv:1705.05165.
\newblock \doarXiv{1705.05165}

\bibitem[{{Carnall} {et~al.}(2019{\natexlab{a}}){Carnall}, {Leja}, {Johnson},
  {McLure}, {Dunlop}, \& {Conroy}}]{car19a}
{Carnall}, A.~C., {Leja}, J., {Johnson}, B.~D., {et~al.} 2019{\natexlab{a}},
  \apj, 873, 44, \dodoi{10.3847/1538-4357/ab04a2}

\bibitem[{{Carnall} {et~al.}(2018){Carnall}, {McLure}, {Dunlop}, \&
  {Dav{\'e}}}]{car18}
{Carnall}, A.~C., {McLure}, R.~J., {Dunlop}, J.~S., \& {Dav{\'e}}, R. 2018,
  \mnras, 480, 4379, \dodoi{10.1093/mnras/sty2169}

\bibitem[{{Carnall} {et~al.}(2019{\natexlab{b}}){Carnall}, {McLure}, {Dunlop},
  {Cullen}, {McLeod}, {Wild}, {Johnson}, {Appleby}, {Dav{\'e}}, {Amorin},
  {Bolzonella}, {Castellano}, {Cimatti}, {Cucciati}, {Gargiulo}, {Garilli},
  {Marchi}, {Pentericci}, {Pozzetti}, {Schreiber}, {Talia}, \&
  {Zamorani}}]{car19b}
{Carnall}, A.~C., {McLure}, R.~J., {Dunlop}, J.~S., {et~al.}
  2019{\natexlab{b}}, \mnras, 490, 417, \dodoi{10.1093/mnras/stz2544}

\bibitem[{{Chabrier}(2003)}]{cha03}
{Chabrier}, G. 2003, \pasp, 115, 763, \dodoi{10.1086/376392}

\bibitem[{{Charlot} \& {Fall}(2000)}]{cf00}
{Charlot}, S., \& {Fall}, S.~M. 2000, \apj, 539, 718, \dodoi{10.1086/309250}

\bibitem[{{Chevallard} \& {Charlot}(2016)}]{che16}
{Chevallard}, J., \& {Charlot}, S. 2016, \mnras, 462, 1415,
  \dodoi{10.1093/mnras/stw1756}

\bibitem[{{Civano} {et~al.}(2016){Civano}, {Marchesi}, {Comastri}, {Urry},
  {Elvis}, {Cappelluti}, {Puccetti}, {Brusa}, {Zamorani}, {Hasinger},
  {Aldcroft}, {Alexander}, {Allevato}, {Brunner}, {Capak}, {Finoguenov},
  {Fiore}, {Fruscione}, {Gilli}, {Glotfelty}, {Griffiths}, {Hao}, {Harrison},
  {Jahnke}, {Kartaltepe}, {Karim}, {LaMassa}, {Lanzuisi}, {Miyaji}, {Ranalli},
  {Salvato}, {Sargent}, {Scoville}, {Schawinski}, {Schinnerer}, {Silverman},
  {Smolcic}, {Stern}, {Toft}, {Trakhtenbrot}, {Treister}, \& {Vignali}}]{civ16}
{Civano}, F., {Marchesi}, S., {Comastri}, A., {et~al.} 2016, \apj, 819, 62,
  \dodoi{10.3847/0004-637X/819/1/62}

\bibitem[{{D'Eugenio} {et~al.}(2020){D'Eugenio}, {van der Wel}, {Wu
  (吳柏锋)}, {Barone}, {van Houdt}, {Bezanson}, {Straatman}, {Pacifici},
  {Muzzin}, {Gallazzi}, {Wild}, {Sobral}, {Bell}, {Zibetti}, {Mowla}, \&
  {Franx}}]{deu20a}
{D'Eugenio}, F., {van der Wel}, A., {Wu (吳柏锋)}, P.-F., {et~al.} 2020,
  \mnras, 497, 389, \dodoi{10.1093/mnras/staa1937}

\bibitem[{{Draine} \& {Li}(2007)}]{dra07}
{Draine}, B.~T., \& {Li}, A. 2007, \apj, 657, 810, \dodoi{10.1086/511055}

\bibitem[{{Dressler} \& {Gunn}(1983)}]{dre83}
{Dressler}, A., \& {Gunn}, J.~E. 1983, \apj, 270, 7, \dodoi{10.1086/161093}

\bibitem[{{Dressler} {et~al.}(1999){Dressler}, {Smail}, {Poggianti}, {Butcher},
  {Couch}, {Ellis}, \& {Oemler}}]{dre99}
{Dressler}, A., {Smail}, I., {Poggianti}, B.~M., {et~al.} 1999, The
  Astrophysical Journal Supplement Series, 122, 51, \dodoi{10.1086/313213}

\bibitem[{{Faber} {et~al.}(2007){Faber}, {Willmer}, {Wolf}, {Koo}, {Weiner},
  {Newman}, {Im}, {Coil}, {Conroy}, {Cooper}, {Davis}, {Finkbeiner}, {Gerke},
  {Gebhardt}, {Groth}, {Guhathakurta}, {Harker}, {Kaiser}, {Kassin},
  {Kleinheinrich}, {Konidaris}, {Kron}, {Lin}, {Luppino}, {Madgwick},
  {Meisenheimer}, {Noeske}, {Phillips}, {Sarajedini}, {Schiavon}, {Simard},
  {Szalay}, {Vogt}, \& {Yan}}]{fab07}
{Faber}, S.~M., {Willmer}, C.~N.~A., {Wolf}, C., {et~al.} 2007, \apj, 665, 265,
  \dodoi{10.1086/519294}

\bibitem[{{Falc{\'o}n-Barroso} {et~al.}(2011){Falc{\'o}n-Barroso},
  {S{\'a}nchez-Bl{\'a}zquez}, {Vazdekis}, {Ricciardelli}, {Cardiel}, {Cenarro},
  {Gorgas}, \& {Peletier}}]{fal11}
{Falc{\'o}n-Barroso}, J., {S{\'a}nchez-Bl{\'a}zquez}, P., {Vazdekis}, A.,
  {et~al.} 2011, \aap, 532, A95, \dodoi{10.1051/0004-6361/201116842}

\bibitem[{{Franzetti} {et~al.}(2007){Franzetti}, {Scodeggio}, {Garilli},
  {Vergani}, {Maccagni}, {Guzzo}, {Tresse}, {Ilbert}, {Lamareille}, {Contini},
  {Le F{\`e}vre}, {Zamorani}, {Brinchmann}, {Charlot}, {Bottini}, {Le Brun},
  {Picat}, {Scaramella}, {Vettolani}, {Zanichelli}, {Adami}, {Arnouts},
  {Bardelli}, {Bolzonella}, {Cappi}, {Ciliegi}, {Foucaud}, {Gavignaud},
  {Iovino}, {McCracken}, {Marano}, {Marinoni}, {Mazure}, {Meneux}, {Merighi},
  {Paltani}, {Pell{\`o}}, {Pollo}, {Pozzetti}, {Radovich}, {Zucca}, {Cucciati},
  \& {Walcher}}]{fra07}
{Franzetti}, P., {Scodeggio}, M., {Garilli}, B., {et~al.} 2007, \aap, 465, 711,
  \dodoi{10.1051/0004-6361:20065942}

\bibitem[{{Girardi} {et~al.}(2000){Girardi}, {Bressan}, {Bertelli}, \&
  {Chiosi}}]{gir00}
{Girardi}, L., {Bressan}, A., {Bertelli}, G., \& {Chiosi}, C. 2000, \aaps, 141,
  371, \dodoi{10.1051/aas:2000126}

\bibitem[{{Gon{\c{c}}alves} {et~al.}(2012){Gon{\c{c}}alves}, {Martin},
  {Men{\'e}ndez-Delmestre}, {Wyder}, \& {Koekemoer}}]{gon12}
{Gon{\c{c}}alves}, T.~S., {Martin}, D.~C., {Men{\'e}ndez-Delmestre}, K.,
  {Wyder}, T.~K., \& {Koekemoer}, A. 2012, \apj, 759, 67,
  \dodoi{10.1088/0004-637X/759/1/67}

\bibitem[{{Goto}(2005)}]{got05}
{Goto}, T. 2005, \mnras, 357, 937, \dodoi{10.1111/j.1365-2966.2005.08701.x}

\bibitem[{{Greene} {et~al.}(2020){Greene}, {Setton}, {Bezanson}, {Suess},
  {Kriek}, {Spilker}, {Goulding}, \& {Feldmann}}]{gre20}
{Greene}, J.~E., {Setton}, D., {Bezanson}, R., {et~al.} 2020, \apjl, 899, L9,
  \dodoi{10.3847/2041-8213/aba534}

\bibitem[{{Hayward} {et~al.}(2014){Hayward}, {Lanz}, {Ashby}, {Fazio},
  {Hernquist}, {Mart{\'\i}nez-Galarza}, {Noeske}, {Smith}, {Wuyts}, \&
  {Zezas}}]{hay14}
{Hayward}, C.~C., {Lanz}, L., {Ashby}, M. L.~N., {et~al.} 2014, \mnras, 445,
  1598, \dodoi{10.1093/mnras/stu1843}

\bibitem[{{Himoto} \& {Kajisawa}(2023)}]{him23}
{Himoto}, K.~G., \& {Kajisawa}, M. 2023, \mnras, 519, 4110,
  \dodoi{10.1093/mnras/stac3687}

\bibitem[{{Ho} {et~al.}(2014){Ho}, {Kewley}, {Dopita}, {Medling}, {Allen},
  {Bland-Hawthorn}, {Bloom}, {Bryant}, {Croom}, {Fogarty}, {Goodwin}, {Green},
  {Konstantopoulos}, {Lawrence}, {L{\'o}pez-S{\'a}nchez}, {Owers}, {Richards},
  \& {Sharp}}]{ho14}
{Ho}, I.~T., {Kewley}, L.~J., {Dopita}, M.~A., {et~al.} 2014, \mnras, 444,
  3894, \dodoi{10.1093/mnras/stu1653}

\bibitem[{{Ho} \& {Keto}(2007)}]{ho07}
{Ho}, L.~C., \& {Keto}, E. 2007, \apj, 658, 314, \dodoi{10.1086/511260}

\bibitem[{{Hopkins} {et~al.}(2006){Hopkins}, {Hernquist}, {Cox}, {Di Matteo},
  {Robertson}, \& {Springel}}]{hop06}
{Hopkins}, P.~F., {Hernquist}, L., {Cox}, T.~J., {et~al.} 2006, The
  Astrophysical Journal Supplement Series, 163, 1, \dodoi{10.1086/499298}

\bibitem[{{Hunter}(2007)}]{hun07}
{Hunter}, J.~D. 2007, Computing in Science and Engineering, 9, 90,
  \dodoi{10.1109/MCSE.2007.55}

\bibitem[{{Ilbert} {et~al.}(2013){Ilbert}, {McCracken}, {Le F{\`e}vre},
  {Capak}, {Dunlop}, {Karim}, {Renzini}, {Caputi}, {Boissier}, {Arnouts},
  {Aussel}, {Comparat}, {Guo}, {Hudelot}, {Kartaltepe}, {Kneib}, {Krogager},
  {Le Floc'h}, {Lilly}, {Mellier}, {Milvang-Jensen}, {Moutard}, {Onodera},
  {Richard}, {Salvato}, {Sanders}, {Scoville}, {Silverman}, {Taniguchi},
  {Tasca}, {Thomas}, {Toft}, {Tresse}, {Vergani}, {Wolk}, \& {Zirm}}]{ilb13}
{Ilbert}, O., {McCracken}, H.~J., {Le F{\`e}vre}, O., {et~al.} 2013, \aap, 556,
  A55, \dodoi{10.1051/0004-6361/201321100}

\bibitem[{{Jin} {et~al.}(2018){Jin}, {Daddi}, {Liu}, {Smol{\v{c}}i{\'c}},
  {Schinnerer}, {Calabr{\`o}}, {Gu}, {Delhaize}, {Delvecchio}, {Gao},
  {Salvato}, {Puglisi}, {Dickinson}, {Bertoldi}, {Sargent}, {Novak}, {Magdis},
  {Aretxaga}, {Wilson}, \& {Capak}}]{jin18}
{Jin}, S., {Daddi}, E., {Liu}, D., {et~al.} 2018, \apj, 864, 56,
  \dodoi{10.3847/1538-4357/aad4af}

\bibitem[{{Juneau} {et~al.}(2014){Juneau}, {Bournaud}, {Charlot}, {Daddi},
  {Elbaz}, {Trump}, {Brinchmann}, {Dickinson}, {Duc}, {Gobat}, {Jean-Baptiste},
  {Le Floc'h}, {Lehnert}, {Pacifici}, {Pannella}, \& {Schreiber}}]{jun14}
{Juneau}, S., {Bournaud}, F., {Charlot}, S., {et~al.} 2014, \apj, 788, 88,
  \dodoi{10.1088/0004-637X/788/1/88}

\bibitem[{{Kennicutt} \& {Evans}(2012)}]{ken12}
{Kennicutt}, R.~C., \& {Evans}, N.~J. 2012, \araa, 50, 531,
  \dodoi{10.1146/annurev-astro-081811-125610}

\bibitem[{{Kewley} {et~al.}(2006){Kewley}, {Groves}, {Kauffmann}, \&
  {Heckman}}]{kew06}
{Kewley}, L.~J., {Groves}, B., {Kauffmann}, G., \& {Heckman}, T. 2006, \mnras,
  372, 961, \dodoi{10.1111/j.1365-2966.2006.10859.x}

\bibitem[{{Kocevski} {et~al.}(2011){Kocevski}, {Lemaux}, {Lubin}, {Gal},
  {McGrath}, {Fassnacht}, {Squires}, {Surace}, \& {Lacy}}]{koc11}
{Kocevski}, D.~D., {Lemaux}, B.~C., {Lubin}, L.~M., {et~al.} 2011, \apj, 736,
  38, \dodoi{10.1088/0004-637X/736/1/38}

\bibitem[{{Kriek} {et~al.}(2011){Kriek}, {van Dokkum}, {Whitaker}, {Labb{\'e}},
  {Franx}, \& {Brammer}}]{kri11}
{Kriek}, M., {van Dokkum}, P.~G., {Whitaker}, K.~E., {et~al.} 2011, \apj, 743,
  168, \dodoi{10.1088/0004-637X/743/2/168}

\bibitem[{{Le Borgne} {et~al.}(2006){Le Borgne}, {Abraham}, {Daniel},
  {McCarthy}, {Glazebrook}, {Savaglio}, {Crampton}, {Juneau}, {Carlberg},
  {Chen}, {Marzke}, {Roth}, {J{\o}rgensen}, \& {Murowinski}}]{leb06}
{Le Borgne}, D., {Abraham}, R., {Daniel}, K., {et~al.} 2006, \apj, 642, 48,
  \dodoi{10.1086/500005}

\bibitem[{{Leja} {et~al.}(2019){Leja}, {Carnall}, {Johnson}, {Conroy}, \&
  {Speagle}}]{lej19a}
{Leja}, J., {Carnall}, A.~C., {Johnson}, B.~D., {Conroy}, C., \& {Speagle},
  J.~S. 2019, \apj, 876, 3, \dodoi{10.3847/1538-4357/ab133c}

\bibitem[{{Lemaux} {et~al.}(2010){Lemaux}, {Lubin}, {Shapley}, {Kocevski},
  {Gal}, \& {Squires}}]{lem10}
{Lemaux}, B.~C., {Lubin}, L.~M., {Shapley}, A., {et~al.} 2010, \apj, 716, 970,
  \dodoi{10.1088/0004-637X/716/2/970}

\bibitem[{{Lemaux} {et~al.}(2017){Lemaux}, {Tomczak}, {Lubin}, {Wu}, {Gal},
  {Rumbaugh}, {Kocevski}, \& {Squires}}]{lem17}
{Lemaux}, B.~C., {Tomczak}, A.~R., {Lubin}, L.~M., {et~al.} 2017, \mnras, 472,
  419, \dodoi{10.1093/mnras/stx1579}

\bibitem[{{Liu} {et~al.}(2019){Liu}, {Schinnerer}, {Groves}, {Magnelli},
  {Lang}, {Leslie}, {Jim{\'e}nez-Andrade}, {Riechers}, {Popping}, {Magdis},
  {Daddi}, {Sargent}, {Gao}, {Fudamoto}, {Oesch}, \& {Bertoldi}}]{liu19}
{Liu}, D., {Schinnerer}, E., {Groves}, B., {et~al.} 2019, \apj, 887, 235,
  \dodoi{10.3847/1538-4357/ab578d}

\bibitem[{{Lotz} {et~al.}(2008){Lotz}, {Jonsson}, {Cox}, \& {Primack}}]{lot08}
{Lotz}, J.~M., {Jonsson}, P., {Cox}, T.~J., \& {Primack}, J.~R. 2008, \mnras,
  391, 1137, \dodoi{10.1111/j.1365-2966.2008.14004.x}

\bibitem[{{Lotz} {et~al.}(2010{\natexlab{a}}){Lotz}, {Jonsson}, {Cox}, \&
  {Primack}}]{lot10a}
---. 2010{\natexlab{a}}, \mnras, 404, 590,
  \dodoi{10.1111/j.1365-2966.2010.16269.x}

\bibitem[{{Lotz} {et~al.}(2010{\natexlab{b}}){Lotz}, {Jonsson}, {Cox}, \&
  {Primack}}]{lot10b}
---. 2010{\natexlab{b}}, \mnras, 404, 575,
  \dodoi{10.1111/j.1365-2966.2010.16268.x}

\bibitem[{{Maltby} {et~al.}(2018){Maltby}, {Almaini}, {Wild}, {Hatch},
  {Hartley}, {Simpson}, {Rowlands}, \& {Socolovsky}}]{mal18}
{Maltby}, D.~T., {Almaini}, O., {Wild}, V., {et~al.} 2018, \mnras, 480, 381,
  \dodoi{10.1093/mnras/sty1794}

\bibitem[{{Marchesi} {et~al.}(2016){Marchesi}, {Civano}, {Elvis}, {Salvato},
  {Brusa}, {Comastri}, {Gilli}, {Hasinger}, {Lanzuisi}, {Miyaji}, {Treister},
  {Urry}, {Vignali}, {Zamorani}, {Allevato}, {Cappelluti}, {Cardamone},
  {Finoguenov}, {Griffiths}, {Karim}, {Laigle}, {LaMassa}, {Jahnke}, {Ranalli},
  {Schawinski}, {Schinnerer}, {Silverman}, {Smolcic}, {Suh}, \&
  {Trakhtenbrot}}]{mar16}
{Marchesi}, S., {Civano}, F., {Elvis}, M., {et~al.} 2016, \apj, 817, 34,
  \dodoi{10.3847/0004-637X/817/1/34}

\bibitem[{{Marigo} \& {Girardi}(2007)}]{mar07}
{Marigo}, P., \& {Girardi}, L. 2007, \aap, 469, 239,
  \dodoi{10.1051/0004-6361:20066772}

\bibitem[{{Maseda} {et~al.}(2021){Maseda}, {van der Wel}, {Franx}, {Bell},
  {Bezanson}, {Muzzin}, {Sobral}, {D'Eugenio}, {Gallazzi}, {de Graaff}, {Leja},
  {Straatman}, {Whitaker}, {Williams}, \& {Wu}}]{mas21}
{Maseda}, M.~V., {van der Wel}, A., {Franx}, M., {et~al.} 2021, \apj, 923, 18,
  \dodoi{10.3847/1538-4357/ac2bfe}

\bibitem[{{Matharu} {et~al.}(2021){Matharu}, {Muzzin}, {Brammer}, {Nelson},
  {Auger}, {Hewett}, {van der Burg}, {Balogh}, {Demarco}, {Marchesini},
  {Noble}, {Rudnick}, {van der Wel}, {Wilson}, \& {Yee}}]{mat21}
{Matharu}, J., {Muzzin}, A., {Brammer}, G.~B., {et~al.} 2021, \apj, 923, 222,
  \dodoi{10.3847/1538-4357/ac26c3}

\bibitem[{{Meusinger} {et~al.}(2017){Meusinger}, {Br{\"u}necke}, {Schalldach},
  \& {in der Au}}]{meu17}
{Meusinger}, H., {Br{\"u}necke}, J., {Schalldach}, P., \& {in der Au}, A. 2017,
  \aap, 597, A134, \dodoi{10.1051/0004-6361/201629139}

\bibitem[{{Muzzin} {et~al.}(2012){Muzzin}, {Wilson}, {Yee}, {Gilbank},
  {Hoekstra}, {Demarco}, {Balogh}, {van Dokkum}, {Franx}, {Ellingson}, {Hicks},
  {Nantais}, {Noble}, {Lacy}, {Lidman}, {Rettura}, {Surace}, \& {Webb}}]{muz12}
{Muzzin}, A., {Wilson}, G., {Yee}, H.~K.~C., {et~al.} 2012, \apj, 746, 188,
  \dodoi{10.1088/0004-637X/746/2/188}

\bibitem[{{Muzzin} {et~al.}(2013){Muzzin}, {Marchesini}, {Stefanon}, {Franx},
  {Milvang-Jensen}, {Dunlop}, {Fynbo}, {Brammer}, {Labb{\'e}}, \& {van
  Dokkum}}]{muz13a}
{Muzzin}, A., {Marchesini}, D., {Stefanon}, M., {et~al.} 2013, The
  Astrophysical Journal Supplement Series, 206, 8,
  \dodoi{10.1088/0067-0049/206/1/8}

\bibitem[{{Nielsen} {et~al.}(2012){Nielsen}, {Ridgway}, {De Propris}, \&
  {Goto}}]{nie12}
{Nielsen}, D.~M., {Ridgway}, S.~E., {De Propris}, R., \& {Goto}, T. 2012,
  \apjl, 761, L16, \dodoi{10.1088/2041-8205/761/2/L16}

\bibitem[{{Poggianti} \& {Wu}(2000)}]{pog00}
{Poggianti}, B.~M., \& {Wu}, H. 2000, \apj, 529, 157, \dodoi{10.1086/308243}

\bibitem[{{Rich} {et~al.}(2011){Rich}, {Kewley}, \& {Dopita}}]{ric11}
{Rich}, J.~A., {Kewley}, L.~J., \& {Dopita}, M.~A. 2011, \apj, 734, 87,
  \dodoi{10.1088/0004-637X/734/2/87}

\bibitem[{{Rowlands} {et~al.}(2018){Rowlands}, {Wild}, {Bourne}, {Bremer},
  {Brough}, {Driver}, {Hopkins}, {Owers}, {Phillipps}, {Pimbblet}, {Sansom},
  {Wang}, {Alpaslan}, {Bland-Hawthorn}, {Colless}, {Holwerda}, \&
  {Taylor}}]{row18}
{Rowlands}, K., {Wild}, V., {Bourne}, N., {et~al.} 2018, \mnras, 473, 1168,
  \dodoi{10.1093/mnras/stx1903}

\bibitem[{{Saintonge} {et~al.}(2011){Saintonge}, {Kauffmann}, {Wang}, {Kramer},
  {Tacconi}, {Buchbender}, {Catinella}, {Graci{\'a}-Carpio}, {Cortese},
  {Fabello}, {Fu}, {Genzel}, {Giovanelli}, {Guo}, {Haynes}, {Heckman},
  {Krumholz}, {Lemonias}, {Li}, {Moran}, {Rodriguez-Fernandez}, {Schiminovich},
  {Schuster}, \& {Sievers}}]{sai11}
{Saintonge}, A., {Kauffmann}, G., {Wang}, J., {et~al.} 2011, \mnras, 415, 61,
  \dodoi{10.1111/j.1365-2966.2011.18823.x}

\bibitem[{{S{\'a}nchez-Bl{\'a}zquez} {et~al.}(2006){S{\'a}nchez-Bl{\'a}zquez},
  {Peletier}, {Jim{\'e}nez-Vicente}, {Cardiel}, {Cenarro},
  {Falc{\'o}n-Barroso}, {Gorgas}, {Selam}, \& {Vazdekis}}]{san06}
{S{\'a}nchez-Bl{\'a}zquez}, P., {Peletier}, R.~F., {Jim{\'e}nez-Vicente}, J.,
  {et~al.} 2006, \mnras, 371, 703, \dodoi{10.1111/j.1365-2966.2006.10699.x}

\bibitem[{{Schawinski} {et~al.}(2014){Schawinski}, {Urry}, {Simmons},
  {Fortson}, {Kaviraj}, {Keel}, {Lintott}, {Masters}, {Nichol}, {Sarzi},
  {Skibba}, {Treister}, {Willett}, {Wong}, \& {Yi}}]{sch14}
{Schawinski}, K., {Urry}, C.~M., {Simmons}, B.~D., {et~al.} 2014, \mnras, 440,
  889, \dodoi{10.1093/mnras/stu327}

\bibitem[{{Scoville} {et~al.}(2007){Scoville}, {Abraham}, {Aussel}, {Barnes},
  {Benson}, {Blain}, {Calzetti}, {Comastri}, {Capak}, {Carilli}, {Carlstrom},
  {Carollo}, {Colbert}, {Daddi}, {Ellis}, {Elvis}, {Ewald}, {Fall},
  {Franceschini}, {Giavalisco}, {Green}, {Griffiths}, {Guzzo}, {Hasinger},
  {Impey}, {Kneib}, {Koda}, {Koekemoer}, {Lefevre}, {Lilly}, {Liu},
  {McCracken}, {Massey}, {Mellier}, {Miyazaki}, {Mobasher}, {Mould}, {Norman},
  {Refregier}, {Renzini}, {Rhodes}, {Rich}, {Sanders}, {Schiminovich},
  {Schinnerer}, {Scodeggio}, {Sheth}, {Shopbell}, {Taniguchi}, {Tyson}, {Urry},
  {Van Waerbeke}, {Vettolani}, {White}, \& {Yan}}]{sco07}
{Scoville}, N., {Abraham}, R.~G., {Aussel}, H., {et~al.} 2007, \apjs, 172, 38,
  \dodoi{10.1086/516580}

\bibitem[{{Setton} {et~al.}(2020){Setton}, {Bezanson}, {Suess}, {Hunt},
  {Greene}, {Kriek}, {Spilker}, {Feldmann}, \& {Narayanan}}]{set20}
{Setton}, D.~J., {Bezanson}, R., {Suess}, K.~A., {et~al.} 2020, \apj, 905, 79,
  \dodoi{10.3847/1538-4357/abc265}

\bibitem[{{Setton} {et~al.}(2022){Setton}, {Verrico}, {Bezanson}, {Greene},
  {Suess}, {Goulding}, {Spilker}, {Kriek}, {Feldmann}, {Narayanan},
  {Hall-Hooper}, \& {Kado-Fong}}]{set22}
{Setton}, D.~J., {Verrico}, M., {Bezanson}, R., {et~al.} 2022, \apj, 931, 51,
  \dodoi{10.3847/1538-4357/ac6096}

\bibitem[{{Smercina} {et~al.}(2018){Smercina}, {Smith}, {Dale}, {French},
  {Croxall}, {Zhukovska}, {Togi}, {Bell}, {Crocker}, {Draine}, {Jarrett},
  {Tremonti}, {Yang}, \& {Zabludoff}}]{sme18}
{Smercina}, A., {Smith}, J.~D.~T., {Dale}, D.~A., {et~al.} 2018, \apj, 855, 51,
  \dodoi{10.3847/1538-4357/aaafcd}

\bibitem[{{Smercina} {et~al.}(2022){Smercina}, {Smith}, {French}, {Bell},
  {Dale}, {Medling}, {Nyland}, {Privon}, {Rowlands}, {Walter}, \&
  {Zabludoff}}]{sme22}
{Smercina}, A., {Smith}, J.-D.~T., {French}, K.~D., {et~al.} 2022, \apj, 929,
  154, \dodoi{10.3847/1538-4357/ac5d5f}

\bibitem[{{Smol{\v{c}}i{\'c}} {et~al.}(2017){Smol{\v{c}}i{\'c}}, {Novak},
  {Bondi}, {Ciliegi}, {Mooley}, {Schinnerer}, {Zamorani}, {Navarrete},
  {Bourke}, {Karim}, {Vardoulaki}, {Leslie}, {Delhaize}, {Carilli}, {Myers},
  {Baran}, {Delvecchio}, {Miettinen}, {Banfield}, {Balokovi{\'c}}, {Bertoldi},
  {Capak}, {Frail}, {Hallinan}, {Hao}, {Herrera Ruiz}, {Horesh}, {Ilbert},
  {Intema}, {Jeli{\'c}}, {Kl{\"o}ckner}, {Krpan}, {Kulkarni}, {McCracken},
  {Laigle}, {Middleberg}, {Murphy}, {Sargent}, {Scoville}, \& {Sheth}}]{smo17}
{Smol{\v{c}}i{\'c}}, V., {Novak}, M., {Bondi}, M., {et~al.} 2017, \aap, 602,
  A1, \dodoi{10.1051/0004-6361/201628704}

\bibitem[{{Snyder} {et~al.}(2011){Snyder}, {Cox}, {Hayward}, {Hernquist}, \&
  {Jonsson}}]{sny11}
{Snyder}, G.~F., {Cox}, T.~J., {Hayward}, C.~C., {Hernquist}, L., \& {Jonsson},
  P. 2011, \apj, 741, 77, \dodoi{10.1088/0004-637X/741/2/77}

\bibitem[{{Somerville} \& {Dav{\'e}}(2015)}]{som15}
{Somerville}, R.~S., \& {Dav{\'e}}, R. 2015, \araa, 53, 51,
  \dodoi{10.1146/annurev-astro-082812-140951}

\bibitem[{{Spilker} {et~al.}(2018){Spilker}, {Bezanson}, {Bari{\v{s}}i{\'c}},
  {Bell}, {Lagos}, {Maseda}, {Muzzin}, {Pacifici}, {Sobral}, {Straatman}, {van
  der Wel}, {van Dokkum}, {Weiner}, {Whitaker}, {Williams}, \& {Wu}}]{spi18}
{Spilker}, J., {Bezanson}, R., {Bari{\v{s}}i{\'c}}, I., {et~al.} 2018, \apj,
  860, 103, \dodoi{10.3847/1538-4357/aac438}

\bibitem[{{Spilker} {et~al.}(2022){Spilker}, {Suess}, {Setton}, {Bezanson},
  {Feldmann}, {Greene}, {Kriek}, {Lower}, {Narayanan}, \& {Verrico}}]{spi22}
{Spilker}, J.~S., {Suess}, K.~A., {Setton}, D.~J., {et~al.} 2022, \apjl, 936,
  L11, \dodoi{10.3847/2041-8213/ac75ea}

\bibitem[{{Springel} {et~al.}(2005){Springel}, {Di Matteo}, \&
  {Hernquist}}]{spr05}
{Springel}, V., {Di Matteo}, T., \& {Hernquist}, L. 2005, \apj, 620, L79,
  \dodoi{10.1086/428772}

\bibitem[{{Stern} {et~al.}(2012){Stern}, {Assef}, {Benford}, {Blain}, {Cutri},
  {Dey}, {Eisenhardt}, {Griffith}, {Jarrett}, {Lake}, {Masci}, {Petty},
  {Stanford}, {Tsai}, {Wright}, {Yan}, {Harrison}, \& {Madsen}}]{ste12}
{Stern}, D., {Assef}, R.~J., {Benford}, D.~J., {et~al.} 2012, \apj, 753, 30,
  \dodoi{10.1088/0004-637X/753/1/30}

\bibitem[{{Straatman} {et~al.}(2018){Straatman}, {van der Wel}, {Bezanson},
  {Pacifici}, {Gallazzi}, {Wu}, {Noeske}, {Bari{\v{s}}i{\'c}}, {Bell},
  {Brammer}, {Calhau}, {Chauke}, {Franx}, {van Houdt}, {Labb{\'e}}, {Maseda},
  {Mu{\~n}oz-Mateos}, {Muzzin}, {van de Sand e}, {Sobral}, \&
  {Spilker}}]{str18}
{Straatman}, C. M.~S., {van der Wel}, A., {Bezanson}, R., {et~al.} 2018, The
  Astrophysical Journal Supplement Series, 239, 27,
  \dodoi{10.3847/1538-4365/aae37a}

\bibitem[{{Strateva} {et~al.}(2001){Strateva}, {Ivezi{\'c}}, {Knapp},
  {Narayanan}, {Strauss}, {Gunn}, {Lupton}, {Schlegel}, {Bahcall}, {Brinkmann},
  {Brunner}, {Budav{\'a}ri}, {Csabai}, {Castander}, {Doi}, {Fukugita},
  {Gy{\'{o}}ry}, {Hamabe}, {Hennessy}, {Ichikawa}, {Kunszt}, {Lamb}, {McKay},
  {Okamura}, {Racusin}, {Sekiguchi}, {Schneider}, {Shimasaku}, \&
  {York}}]{str01}
{Strateva}, I., {Ivezi{\'c}}, {\v{Z}}., {Knapp}, G.~R., {et~al.} 2001, \aj,
  122, 1861, \dodoi{10.1086/323301}

\bibitem[{{Suess} {et~al.}(2017){Suess}, {Bezanson}, {Spilker}, {Kriek},
  {Greene}, {Feldmann}, {Hunt}, \& {Narayanan}}]{sue17}
{Suess}, K.~A., {Bezanson}, R., {Spilker}, J.~S., {et~al.} 2017, \apjl, 846,
  L14, \dodoi{10.3847/2041-8213/aa85dc}

\bibitem[{{Suess} {et~al.}(2022){Suess}, {Kriek}, {Bezanson}, {Greene},
  {Setton}, {Spilker}, {Feldmann}, {Goulding}, {Johnson}, {Leja}, {Narayanan},
  {Hall-Hooper}, {Hunt}, {Lower}, \& {Verrico}}]{sue22}
{Suess}, K.~A., {Kriek}, M., {Bezanson}, R., {et~al.} 2022, \apj, 926, 89,
  \dodoi{10.3847/1538-4357/ac404a}

\bibitem[{{Tacconi} {et~al.}(2013){Tacconi}, {Neri}, {Genzel}, {Combes},
  {Bolatto}, {Cooper}, {Wuyts}, {Bournaud}, {Burkert}, {Comerford}, {Cox},
  {Davis}, {F{\"o}rster Schreiber}, {Garc{\'{\i}}a-Burillo}, {Gracia-Carpio},
  {Lutz}, {Naab}, {Newman}, {Omont}, {Saintonge}, {Shapiro Griffin}, {Shapley},
  {Sternberg}, \& {Weiner}}]{tac13}
{Tacconi}, L.~J., {Neri}, R., {Genzel}, R., {et~al.} 2013, \apj, 768, 74,
  \dodoi{10.1088/0004-637X/768/1/74}

\bibitem[{{The Astropy Collaboration} {et~al.}(2018){The Astropy
  Collaboration}, {Price-Whelan}, {Sip{\H o}cz}, {G{\"u}nther}, {Lim},
  {Crawford}, {Conseil}, {Shupe}, {Craig}, {Dencheva}, {Ginsburg},
  {VanderPlas}, {Bradley}, {P{\'e}rez-Su{\'a}rez}, {de Val-Borro}, {Aldcroft},
  {Cruz}, {Robitaille}, {Tollerud}, {Ardelean}, {Babej}, {Bachetti}, {Bakanov},
  {Bamford}, {Barentsen}, {Barmby}, {Baumbach}, {Berry}, {Biscani}, {Boquien},
  {Bostroem}, {Bouma}, {Brammer}, {Bray}, {Breytenbach}, {Buddelmeijer},
  {Burke}, {Calderone}, {Cano Rodr{\'{\i}}guez}, {Cara}, {Cardoso},
  {Cheedella}, {Copin}, {Crichton}, {D{\'A}vella}, {Deil}, {Depagne},
  {Dietrich}, {Donath}, {Droettboom}, {Earl}, {Erben}, {Fabbro}, {Ferreira},
  {Finethy}, {Fox}, {Garrison}, {Gibbons}, {Goldstein}, {Gommers}, {Greco},
  {Greenfield}, {Groener}, {Grollier}, {Hagen}, {Hirst}, {Homeier}, {Horton},
  {Hosseinzadeh}, {Hu}, {Hunkeler}, {Ivezi{\'c}}, {Jain}, {Jenness}, {Kanarek},
  {Kendrew}, {Kern}, {Kerzendorf}, {Khvalko}, {King}, {Kirkby}, {Kulkarni},
  {Kumar}, {Lee}, {Lenz}, {Littlefair}, {Ma}, {Macleod}, {Mastropietro},
  {McCully}, {Montagnac}, {Morris}, {Mueller}, {Mumford}, {Muna}, {Murphy},
  {Nelson}, {Nguyen}, {Ninan}, {N{\"o}the}, {Ogaz}, {Oh}, {Parejko}, {Parley},
  {Pascual}, {Patil}, {Patil}, {Plunkett}, {Prochaska}, {Rastogi}, {Reddy
  Janga}, {Sabater}, {Sakurikar}, {Seifert}, {Sherbert}, {Sherwood-Taylor},
  {Shih}, {Sick}, {Silbiger}, {Singanamalla}, {Singer}, {Sladen}, {Sooley},
  {Sornarajah}, {Streicher}, {Teuben}, {Thomas}, {Tremblay}, {Turner},
  {Terr{\'o}n}, {van Kerkwijk}, {de la Vega}, {Watkins}, {Weaver}, {Whitmore},
  {Woillez}, \& {Zabalza}}]{astropy18}
{The Astropy Collaboration}, {Price-Whelan}, A.~M., {Sip{\H o}cz}, B.~M.,
  {et~al.} 2018, ArXiv e-prints.
\newblock \doarXiv{1801.02634}

\bibitem[{{Tremonti} {et~al.}(2007){Tremonti}, {Moustakas}, \&
  {Diamond-Stanic}}]{tre07}
{Tremonti}, C.~A., {Moustakas}, J., \& {Diamond-Stanic}, A.~M. 2007, \apjl,
  663, L77, \dodoi{10.1086/520083}

\bibitem[{{Tremonti} {et~al.}(2004){Tremonti}, {Heckman}, {Kauffmann},
  {Brinchmann}, {Charlot}, {White}, {Seibert}, {Peng}, {Schlegel}, {Uomoto},
  {Fukugita}, \& {Brinkmann}}]{tre04}
{Tremonti}, C.~A., {Heckman}, T.~M., {Kauffmann}, G., {et~al.} 2004, \apj, 613,
  898, \dodoi{10.1086/423264}

\bibitem[{{van der Wel} {et~al.}(2016){van der Wel}, {Noeske}, {Bezanson},
  {Pacifici}, {Gallazzi}, {Franx}, {Mu{\~n}oz-Mateos}, {Bell}, {Brammer},
  {Charlot}, {Chauk{\'e}}, {Labb{\'e}}, {Maseda}, {Muzzin}, {Rix}, {Sobral},
  {van de Sande}, {van Dokkum}, {Wild}, \& {Wolf}}]{vdw16}
{van der Wel}, A., {Noeske}, K., {Bezanson}, R., {et~al.} 2016, \apjs, 223, 29,
  \dodoi{10.3847/0067-0049/223/2/29}

\bibitem[{{van der Wel} {et~al.}(2021){van der Wel}, {Bezanson}, {D'Eugenio},
  {Straatman}, {Franx}, {van Houdt}, {Maseda}, {Gallazzi}, {Wu}, {Pacifici},
  {Barisic}, {Brammer}, {Munoz-Mateos}, {Vervalcke}, {Zibetti}, {Sobral}, {de
  Graaff}, {Calhau}, {Kaushal}, {Muzzin}, {Bell}, \& {van Dokkum}}]{vdw21}
{van der Wel}, A., {Bezanson}, R., {D'Eugenio}, F., {et~al.} 2021, \apjs, 256,
  44, \dodoi{10.3847/1538-4365/ac1356}

\bibitem[{{Vazdekis}(1999)}]{vaz99}
{Vazdekis}, A. 1999, \apj, 513, 224, \dodoi{10.1086/306843}

\bibitem[{{Vergani} {et~al.}(2010){Vergani}, {Zamorani}, {Lilly}, {Lamareille},
  {Halliday}, {Scodeggio}, {Vignali}, {Ciliegi}, {Bolzonella}, {Bondi},
  {Kova{\v{c}}}, {Knobel}, {Zucca}, {Caputi}, {Pozzetti}, {Bardelli},
  {Mignoli}, {Iovino}, {Carollo}, {Contini}, {Kneib}, {Le F{\`e}vre},
  {Mainieri}, {Renzini}, {Bongiorno}, {Coppa}, {Cucciati}, {de la Torre}, {de
  Ravel}, {Franzetti}, {Garilli}, {Kampczyk}, {Le Borgne}, {Le Brun}, {Maier},
  {Pello}, {Peng}, {Perez Montero}, {Ricciardelli}, {Silverman}, {Tanaka},
  {Tasca}, {Tresse}, {Abbas}, {Bottini}, {Cappi}, {Cassata}, {Cimatti},
  {Guzzo}, {Koekemoer}, {Leauthaud}, {Maccagni}, {Marinoni}, {McCracken},
  {Memeo}, {Meneux}, {Oesch}, {Porciani}, {Scaramella}, {Capak}, {Sanders},
  {Scoville}, \& {Taniguchi}}]{ver10}
{Vergani}, D., {Zamorani}, G., {Lilly}, S., {et~al.} 2010, \aap, 509, A42,
  \dodoi{10.1051/0004-6361/200912802}

\bibitem[{{Verrico} {et~al.}(2022){Verrico}, {Setton}, {Bezanson}, {Greene},
  {Suess}, {Goulding}, {Spilker}, {Kriek}, {Feldmann}, {Narayanan}, {Donofrio},
  \& {Khullar}}]{ver22}
{Verrico}, M., {Setton}, D.~J., {Bezanson}, R., {et~al.} 2022, arXiv e-prints,
  arXiv:2211.16532, \dodoi{10.48550/arXiv.2211.16532}

\bibitem[{{Wang} {et~al.}(2022){Wang}, {Magnelli}, {Schinnerer}, {Liu},
  {Modak}, {Jim{\'e}nez-Andrade}, {Karoumpis}, {Kokorev}, \&
  {Bertoldi}}]{wan22}
{Wang}, T.-M., {Magnelli}, B., {Schinnerer}, E., {et~al.} 2022, \aap, 660,
  A142, \dodoi{10.1051/0004-6361/202142299}

\bibitem[{{Weaver} {et~al.}(2022){Weaver}, {Kauffmann}, {Ilbert}, {McCracken},
  {Moneti}, {Toft}, {Brammer}, {Shuntov}, {Davidzon}, {Hsieh}, {Laigle},
  {Anastasiou}, {Jespersen}, {Vinther}, {Capak}, {Casey}, {McPartland},
  {Milvang-Jensen}, {Mobasher}, {Sanders}, {Zalesky}, {Arnouts}, {Aussel},
  {Dunlop}, {Faisst}, {Franx}, {Furtak}, {Fynbo}, {Gould}, {Greve}, {Gwyn},
  {Kartaltepe}, {Kashino}, {Koekemoer}, {Kokorev}, {Le F{\`e}vre}, {Lilly},
  {Masters}, {Magdis}, {Mehta}, {Peng}, {Riechers}, {Salvato}, {Sawicki},
  {Scarlata}, {Scoville}, {Shirley}, {Silverman}, {Sneppen}, {Smolc̆i{\'c}},
  {Steinhardt}, {Stern}, {Tanaka}, {Taniguchi}, {Teplitz}, {Vaccari}, {Wang},
  \& {Zamorani}}]{wea22}
{Weaver}, J.~R., {Kauffmann}, O.~B., {Ilbert}, O., {et~al.} 2022, \apjs, 258,
  11, \dodoi{10.3847/1538-4365/ac3078}

\bibitem[{{Whitaker} {et~al.}(2012){Whitaker}, {van Dokkum}, {Brammer}, \&
  {Franx}}]{whi12}
{Whitaker}, K.~E., {van Dokkum}, P.~G., {Brammer}, G., \& {Franx}, M. 2012,
  \apjl, 754, L29, \dodoi{10.1088/2041-8205/754/2/L29}

\bibitem[{{Wild} {et~al.}(2016){Wild}, {Almaini}, {Dunlop}, {Simpson},
  {Rowlands}, {Bowler}, {Maltby}, \& {McLure}}]{wil16}
{Wild}, V., {Almaini}, O., {Dunlop}, J., {et~al.} 2016, \mnras, 463, 832,
  \dodoi{10.1093/mnras/stw1996}

\bibitem[{{Wild} {et~al.}(2009){Wild}, {Walcher}, {Johansson}, {Tresse},
  {Charlot}, {Pollo}, {Le F{\`e}vre}, \& {de Ravel}}]{wil09}
{Wild}, V., {Walcher}, C.~J., {Johansson}, P.~H., {et~al.} 2009, \mnras, 395,
  144, \dodoi{10.1111/j.1365-2966.2009.14537.x}

\bibitem[{{Wild} {et~al.}(2020){Wild}, {Taj Aldeen}, {Carnall}, {Maltby},
  {Almaini}, {Werle}, {Wilkinson}, {Rowlands}, {Bolzonella}, {Castellano},
  {Gargiulo}, {McLure}, {Pentericci}, \& {Pozzetti}}]{wil20}
{Wild}, V., {Taj Aldeen}, L., {Carnall}, A., {et~al.} 2020, \mnras, 494, 529,
  \dodoi{10.1093/mnras/staa674}

\bibitem[{{Willmer} {et~al.}(2006){Willmer}, {Faber}, {Koo}, {Weiner},
  {Newman}, {Coil}, {Connolly}, {Conroy}, {Cooper}, {Davis}, {Finkbeiner},
  {Gerke}, {Guhathakurta}, {Harker}, {Kaiser}, {Kassin}, {Konidaris}, {Lin},
  {Luppino}, {Madgwick}, {Noeske}, {Phillips}, \& {Yan}}]{wil06}
{Willmer}, C.~N.~A., {Faber}, S.~M., {Koo}, D.~C., {et~al.} 2006, \apj, 647,
  853, \dodoi{10.1086/505455}

\bibitem[{{Worthey}(1994)}]{wor94}
{Worthey}, G. 1994, \apjs, 95, 107, \dodoi{10.1086/192096}

\bibitem[{{Wu}(2021)}]{wu21a}
{Wu}, P.-F. 2021, \apj, 913, 44, \dodoi{10.3847/1538-4357/abf493}

\bibitem[{{Wu} {et~al.}(2014){Wu}, {Gal}, {Lemaux}, {Kocevski}, {Lubin},
  {Rumbaugh}, \& {Squires}}]{wu14}
{Wu}, P.-F., {Gal}, R.~R., {Lemaux}, B.~C., {et~al.} 2014, \apj, 792, 16,
  \dodoi{10.1088/0004-637X/792/1/16}

\bibitem[{{Wu} {et~al.}(2018){Wu}, {van der Wel}, {Bezanson}, {Gallazzi},
  {Pacifici}, {Straatman}, {Bari{\v{s}}i{\'c}}, {Bell}, {Chauke}, {van Houdt},
  {Franx}, {Muzzin}, {Sobral}, \& {Wild}}]{wu18a}
{Wu}, P.-F., {van der Wel}, A., {Bezanson}, R., {et~al.} 2018, \apj, 868, 37,
  \dodoi{10.3847/1538-4357/aae822}

\bibitem[{{Wu} {et~al.}(2020){Wu}, {van der Wel}, {Bezanson}, {Gallazzi},
  {Pacifici}, {Straatman}, {Bari{\v{s}}i{\'c}}, {Bell}, {Chauke},
  {D{\textquoteright}Eugenio}, {Franx}, {Muzzin}, {Sobral}, \& {van
  Houdt}}]{wu20}
---. 2020, \apj, 888, 77, \dodoi{10.3847/1538-4357/ab5fd9}

\bibitem[{{Yesuf} {et~al.}(2014){Yesuf}, {Faber}, {Trump}, {Koo}, {Fang},
  {Liu}, {Wild}, \& {Hayward}}]{yes14}
{Yesuf}, H.~M., {Faber}, S.~M., {Trump}, J.~R., {et~al.} 2014, \apj, 792, 84,
  \dodoi{10.1088/0004-637X/792/2/84}

\bibitem[{{Zabludoff} {et~al.}(1996){Zabludoff}, {Zaritsky}, {Lin}, {Tucker},
  {Hashimoto}, {Shectman}, {Oemler}, \& {Kirshner}}]{zab96}
{Zabludoff}, A.~I., {Zaritsky}, D., {Lin}, H., {et~al.} 1996, \apj, 466, 104,
  \dodoi{10.1086/177495}

\bibitem[{{Zanella} {et~al.}(2023){Zanella}, {Valentino}, {Gallazzi}, {Belli},
  {Magdis}, \& {Bolamperti}}]{zan23}
{Zanella}, A., {Valentino}, F., {Gallazzi}, A., {et~al.} 2023, \mnras,
  \dodoi{10.1093/mnras/stad1821}

\bibitem[{{Zwaan} {et~al.}(2013){Zwaan}, {Kuntschner}, {Pracy}, \&
  {Couch}}]{zwa13}
{Zwaan}, M.~A., {Kuntschner}, H., {Pracy}, M.~B., \& {Couch}, W.~J. 2013,
  \mnras, 432, 492, \dodoi{10.1093/mnras/stt496}

\end{thebibliography}
\bibliographystyle{aasjournal}

\end{document}